\pdfoutput=1   
\documentclass[usenatbib,usegraphicx,fleqn]{ctextemp_mn2e}
\usepackage{amsmath,amssymb,bm,color,multirow,multicol}
\usepackage{charter}   

\newcommand{\e}[1]{\times 10^{#1}}

\usepackage{ulem}

\title[MRI-driven Accretion onto Magnetized stars:]{MRI-driven Accretion onto Magnetized stars: Axisymmetric MHD Simulations}

\author[M. M. Romanova et al.]
{M. M. Romanova,$^1$\thanks{e-mail:romanova@astro.cornell.edu}, G.
V. Ustyugova,$^2$\thanks{e-mail:ustyugg@rambler.ru},
A. V. Koldoba$^2$, R. V. E. Lovelace $^{1,3}$\\
$^1$ Department of Astronomy, Cornell University, Ithaca, NY 14853-6801, USA\\
$^2$ Keldysh Institute for Applied Mathematics, Moscow, Russia\\
$^3$ Department of Applied and Engineering Physics, Cornell
University, Ithaca, NY 14853-6801}

\begin{document}

\maketitle

\begin{abstract}

\noindent We present the first results of a global axisymmetric
simulation of accretion onto rotating \textit{magnetized stars}
from a turbulent accretion disk, where the turbulence is driven by
the magnetorotational instability (MRI). Long-lasting accretion is
observed  for several thousand rotation periods of the inner
part of the disc.  The angular momentum is transported outward by
the magnetic stress of the turbulent flow with a rate
corresponding to a Shakura-Sunyaev viscosity parameter
$\alpha\approx 0.01-0.04$.
  Close to the star the disk is stopped by the magnetic pressure of the magnetosphere.
The subsequent evolution depends on the orientation of the
poloidal field in the disk relative to that of the star at the
disk-magnetosphere boundary. If fields have the same polarity,
then the magnetic flux is accumulated at the boundary and blocks
the accretion which leads to the accumulation of matter at the
boundary. Subsequently, this matter accretes to the star in
outburst before accumulating again. Hence, the cycling, `bursty'
accretion is observed. The magnetic stress is enhanced at the
boundary, leading to the enhanced accretion rate.  If the disc and
stellar fields have opposite polarity, then the field reconnection
enhances the penetration of the disk matter towards the deeper
field lines of the magnetosphere. However, the magnetic stress at
the boundary is lower due to the field reconnection. This
decreases the accretion rate and leads to smoother accretion at a
lower rate. Test simulations show that  in the case of higher
accretion rate  corresponding to $\alpha=0.05-0.1$, accretion is
bursty in cases of both polarities. On the other hand, at much
lower accretion rates corresponding to $\alpha < 0.01$, accretion
is not bursty in any of these cases. We conclude that the
episodic, bursty accretion is expected during periods of higher
accretion rates in the disc, and in some cases it may alternate
between bursty and smooth accretion, if the disk brings the
poloidal field of alternating polarity. We find that a rotating,
magnetically-dominated corona forms above and below the disk, and
that  it slowly expands outward, driven by the magnetic force.

\end{abstract}


\begin{figure*}
\begin{center}
\includegraphics[width=18.0cm]{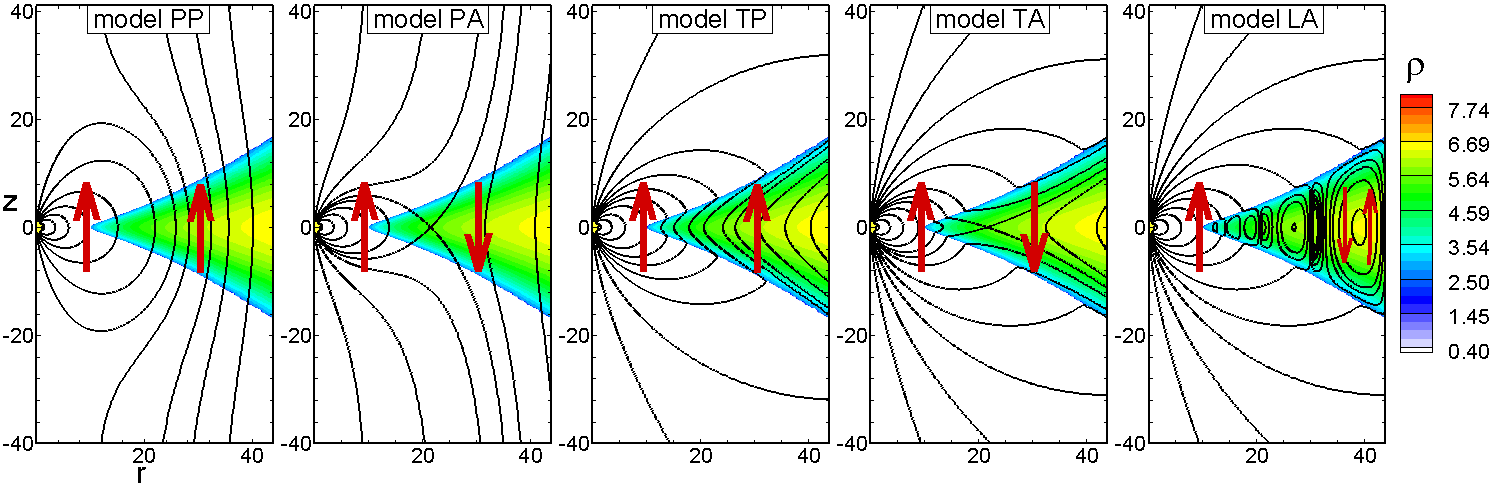}
\caption{\label{init-5} Initial distribution of the magnetic field
used in different models. The dipole magnetic field of the star is
fixed at $\widetilde{\mu}=20$, but the field distribution in the
disc is different. From left to right: the field in the disc is
vertical (parallel to the $z-$axis) and has the same direction as
the magnetic field of the star at the disc-magnetosphere boundary
(model VP20); same but for the opposite direction of the field in
the disc (antiparallel case, model VA20); tapered field with
$z-$component parallel to the field of the star (model TP20);
tapered field with antiparallel field (model TA20); loops in the
disc where the inner loop field is parallel to the star's field
(model LP20).}

\end{center}
\end{figure*}

\begin{figure}
\begin{center}
\includegraphics[width=8cm]{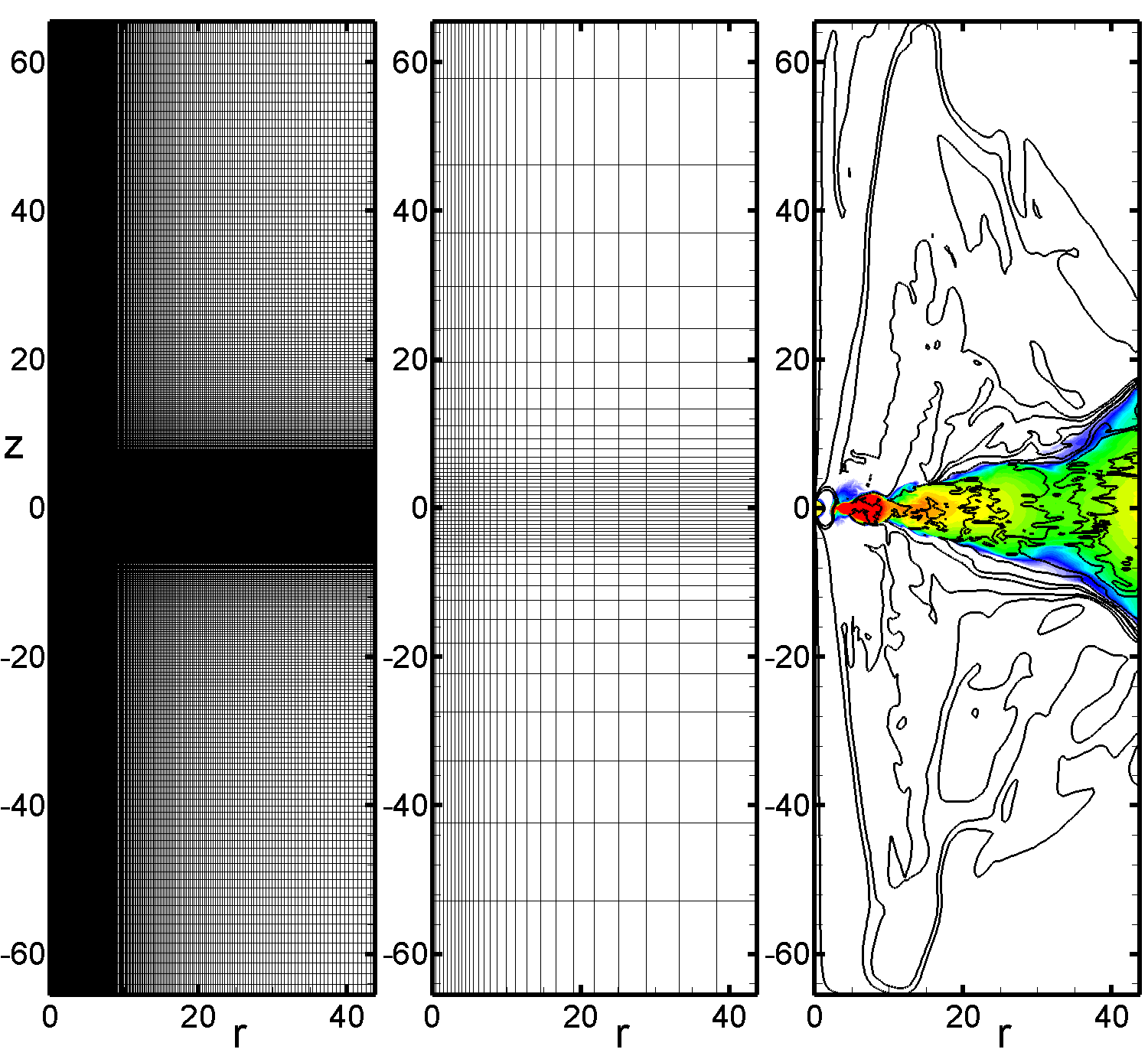}
\caption{\label{grids-3} \textit{Left panel:} The grid used in
simulations; \textit{Middle panel:} same grid, but rarefied by a
factor of 10. \textit{Right panel:} Sample simulation run shows
the density distribution and selected field lines at the end of a
long simulation run in  model \textit{TP20} at $t=3000$.}
\end{center}
\end{figure}

\section{Introduction}

The magnetorotational instability (MRI)  is the likely origin of
turbulent stress in  accretion disks around black holes and stars
(e.g., \citealt{balb91}; \citealt{balb98}). MRI-driven
accretion has been observed in a number of local and global
axisymmetric and 3D MHD simulations (e.g., \citealt{hawl95};
\citealt{ston96}; \citealt{gamm98}; \citealt{ston01};
\citealt{hawl00}; \citealt{hawl01}; \citealt{hawl02};
\citealt{beck09}). In these simulations the central object is
\textit{a black hole}. However, there are many stars which have
dynamically important magnetic fields. These include young,
Classical T Tauri stars (hereafter CTTSs; e.g., \citealt{bouv05}),
some types of white dwarfs, and neutron stars in binary systems
(e.g., \citealt{warn04}; \citealt{vander00}). In these stars, the
magnetic field  is strong enough to stop the disk at radii larger
than the radius of the star. Many observational properties are
determined by the processes at the disk-magnetosphere boundary.

The disk-magnetosphere interaction in the case of laminar
(non-turbulent) flow in the disk has been investigated earlier in
both axisymmetric (e.g., \citealt{mill97}; \citealt{good97};
\citealt{good99}; \citealt{roma02}; \citealt{bess08}) and global
3D MHD simulations (\citealt{roma03}; \citealt{roma04a};
\citealt{kulk05}; \citealt{long08}). The laminar flow in the disc
is  modeled  with the  $\alpha-$viscosity prescription
(\citealt{shak73}). Similarly, the $\alpha-$type magnetic
diffusivity has been included in a number of studies. Heretofore,
accretion due to turbulent, MRI-driven discs has not been studied.

Here, we present the results of the first axisymmetric simulations of
MRI-driven accretion onto magnetized stars. In
these simulations we were able to use a very high grid resolution
and perform very long runs. Results for global 3D simulations will
be discussed in a separate paper \citep{roma11}. Accretion onto rapidly rotating
stars will be described in \citet{usty11}. Here, we consider the
case of slowly rotating stars.


\begin{table}
\centering
\begin{tabular}{llllll}
\hline
Model ~~   & config. & direction    & $B_d$  &$\tilde{\mu}$ &  $t/P_{mid}$ \\
\hline
VP30       & vertical      & parallel      & +0.002  & 30  &19    \\
VA30       & vertical      & antipar.   & -- 0.002 & 30  & 14\\
TP30       & tapered       & parallel       & +0.002& 30  &19\\
TA30       & tapered       & antipar. & -- 0.002 & 30  & 14\\
\hline
VP20       & vertical      & parallel     & +0.002  & 20  & 13 \\
VA20       & vertical      & antipar. & -- 0.002 & 20   & 14\\
TP20       & tapered       & parallel      & +0.002 & 20  & 30 \\
TA20       & tapered       & antipar. & -- 0.002 & 20 & 17\\
LP20       & loops         & parallel       & +0.002& 20  & 13 \\
LA20       & loops         & antipar. & -- 0.002 & 20 & 34\\
\hline
TP20+B001     & tapered       & parallel      & +0.001 & 20  & 12 \\
TA20-B005     & tapered       & antipar. & -- 0.005  & 20  & 33 \\
\hline
TP10       & tapered       & parallel      & +0.002 & 10  & 14\\
TA10       & tapered       & antipar.  & -- 0.002 & 10  & 16\\
\hline
TP5        & tapered       & parallel      & +0.002 & 5  & 23\\
TA5        & tapered       & antipar.  & -- 0.002 & 5 & 13\\
\hline
TP2        & tapered       & parallel      & +0.002 & 2  & 7\\
TA2        & tapered       & antipar.  & -- 0.002 & 2  & 13\\
\hline
\end{tabular}
\caption{The Table describes the main simulation models, where the
first letter in the model name stands for the field configuration:
`V' - vertical field, `T' - tapered field, `L' - looped field. The
second letter stands for the mutual directions of the magnetic
fields of the disc and the star: `P' - parallel and `A' -
antiparallel. The number at the end of the name shows the value of
the dimensionless magnetic moment $\tilde{\mu}$. The Table also shows
the $z-$component of the seed magnetic field in the disc,  $B_d$,
and the duration of simulations in periods of rotation
 $P_{\rm mid}$ at the grid center, $r=22$ ($P_{\rm mid}=103 P_0$).} \label{tab:models}
\end{table}

\section{Theoretical background}
\label{sec:theory}

 The magnetorotational instability (MRI)
arises under conditions where a weak magnetic field threads the
disc. Below, we briefly summarize the condition for the onset of
the MRI instability (Balbus \& Hawley 1991) for a simple case
where
 an axial magnetic field $B_0\hat{\bf z}$ threads a
Keplerian disc which rotates with  angular rate
 $\Omega=(GM/r^3)^{1/2}$ where $M$ is the mass of
 the central object.
      For axisymmetric perturbations of the disc with $\delta{\bf v}=
[\delta v_r(z,t), \delta v_\phi(z,t),0]$ and $\delta {\bf
B}=[\delta B_r(z,t),\delta B_\phi(z,t),0] $ and for perturbations
proportional to $\exp(ik_z z-i\omega t)$, one finds the dispersion
relation
\begin{equation}
\omega_\pm^2 = (k_z v_A)^2 +{1\over 2}\kappa_r^2 \pm \left[{1\over
4}\kappa_r^4+4(k_z v_A\Omega)^2\right]^{1/2}~,
\end{equation}
 where $v_A \equiv B_0/\sqrt{4\pi \rho}$
 is the Alfv\'en velocity and $\kappa_r \equiv [4\Omega^2 +2r\Omega
d\Omega/dr]^{1/2}$ is the radial epicyclic frequency of the disc.
In order for the perturbation to fit within the vertical extent of
the disc  one needs $k_z h \gtrsim 1$, where $h = c_s/\Omega$ is
the half-thickness of the disc and $c_s$ is the isothermal sound
speed in the disc.  For most conditions the disc is thin with $ h
\ll r$, or $c_s \ll r\Omega$.

Evidently, there can be instability if $\omega_-^2 <0$ which
happens if $(kv_A)^2 < -2r\Omega d\Omega/dr$.  For a Keplerian
disc this corresponds to $(kv_A)^2  <3\Omega^2$.
    Therefore, the above-mentioned condition that $k_z h
\gtrsim 1$ implies that the instability occurs only for $ v_A <
c_s$,  or
\begin{equation}
\beta \equiv {2 c_s^2 \over v_A^2} >1~.
\end{equation}
Note that $\beta$ is based on the {\it initial} vertical magnetic
field, $B_0$.   As a result of the instability the magnetic field
may grow to values much larger than $B_0$.
    The maximum value of the growth
rate  is $\Im(\omega)_{\rm max}=3\Omega/4$, and it occurs for
$k_{\rm max}=(15/16)^{1/2}\Omega/v_A$.    For $\beta <1$ the
perturbation does not fit inside the disc and there is stability.
   As $\beta$ increases from unity (weaker magnetic field) the
maximum growth rate stays the same but the wavelength of the
perturbation gets shorter ($\propto \beta^{-1/2}$).
    Of course, for sufficiently small wavelengths the damping
due to numerical viscosity  ($\sim \nu_{\rm num}k^2$) will be
larger than the MRI growth rate. To generate MRI-driven accretion
in numerical simulations, one should have large enough $\beta$ in
the disc so as to have instability, and at the same time high enough
grid resolution to avoid the damping of small wavelengths.

\begin{figure*}
\begin{center}
\includegraphics[width=12cm]{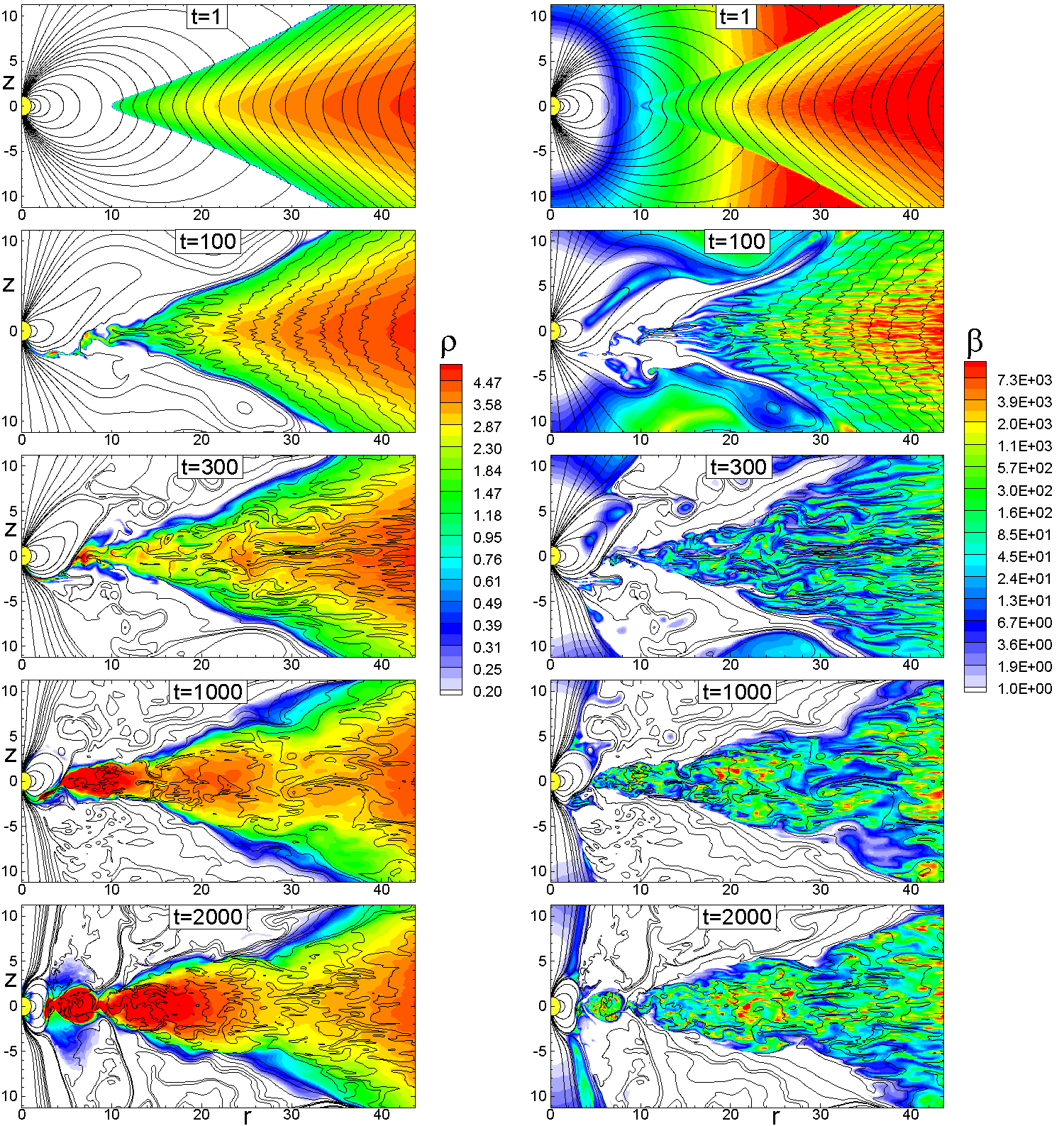}
\caption{\label{beta-rho} MRI-driven accretion at different times
for model TP30. {\it Left panels:} The background shows the
density and the lines are poloidal field lines.  {\it Right
panels:} The background shows the plasma parameter $\beta=8\pi
P_m/B^2$, where $P_m$ is matter pressure.}
\end{center}
\end{figure*}

\section{Simulations of MRI instability}

\subsection{The problem setup}

We investigate MRI-driven accretion onto a rotating star with a
dipole magnetic field.  The field lines are frozen into the
stellar surface, and initially, they thread the entire simulation
region, including the disc. The disc is dense and cold while the
corona is $10^3$ times hotter and $10^3$ times less dense. The
disc rotates with  the Keplerian velocity at the beginning of the
simulation. The corona also rotates with Keplerian velocities at
cylindrical radii corresponding to the rotation of the disc. This
helps to avoid the build-up of an initial discontinuity of the
magnetic field at the disc-magnetosphere boundary. We derive the
initial distribution of density and pressure in the disc and
corona by balancing the gravitational, centrifugal and pressure
gradient forces (see Sec. \ref{sec:app-initial-cond} for details).
The initial inner disc radius is set to $r=10$ (in units of
stellar radius, $R_0$; see Sec. \ref{sec:app-initial-cond} for the
full description of reference units).  The star rotates slowly
with an angular velocity corresponding to a corotation radius
($r_{\rm cor}=(GM/\Omega_*^2)^{1/3}$) of $10$ (in dimensionless
units).

A small seed poloidal field with $z-$component $B_d$ has been
added to the disc, which is necessary for the initial excitation
of the MRI-driven turbulence. We considered three types of initial
configurations:
\begin{itemize}

\item \textit{Vertical field}(`V-'type models): The field is
vertical (that is, parallel to the $z-$axis), has a value of $B_d$
and is homogeneous in space. The alignment of the field may be
parallel or antiparallel to the star's magnetic field at the
disc-magnetosphere boundary (see two left panels in Fig.
\ref{init-5}).

\item \textit{Tapered field} (`T-'type models): An initially
vertical field tapered in such a way that the magnetic flux is
restricted to lie within the disc . The magnetic flux inside a
disc with half-thickness $h$ (at $|z| \le h$) is determined by:

\begin{equation}
\Psi=\frac{B_d r^2}{2} \cos \bigg( \pi \frac{z}{2h} \bigg), ~~~~
h=\sqrt{ \bigg(\frac{GM}{\Phi_c(r)-E}\bigg)^2 - r^2},
\label{eq:tapered}
\end{equation}

\noindent where  $\Phi_c(r)=\kappa GM/r$, $E$ is the constant of
integration in the initial equilibrium equation
 (see Sec. \ref{sec:app-initial-cond}).
Fig. \ref{init-5} (3rd and 4th panels from the left) shows cases
of parallel (model `TP') and antiparallel (model `TA') fields.

\item \textit{Looped field} (`L-'type models): The magnetic flux
inside a disc with half-thickness $h$  is determined by:

\begin{equation}
\Psi =\frac{B_d r^2}{2} \cos \bigg(a \pi \frac{z}{2 h}\bigg) \sin
\bigg(b \pi \frac{r-r_d}{r_{out}-r_d}\bigg)~,~  \label{eq:loops}
\end{equation}

\noindent where parameters  $a$ and $b$ determine the number of
loops in the vertical and horizontal directions,  $r_{out}$ is the
radius of the external boundary, and $r_d$ is the inner disc
radius.

We experimented with different numbers of loops with values of
$Nr_{\rm loops}\times Nz_{\rm loops}$ ranging from $3\times 1$ to
$7\times 5$, and concluded that the case with the smaller number
of loops ($3\times 1$) is sufficient for exciting the turbulence,
while the cases with a larger number of loops do not have
advantages. The total poloidal flux is not zero, and hence we have
two cases where the averaged field in the disc is parallel (model
`LP') or antiparallel (model `LA') to the field of the star (see
an example for `LP-'type configuration at the right panel of Fig.
\ref{init-5}).


\end{itemize}

We solve the ideal MHD equations (see Sec.
\ref{sec:app-equations}), given axisymmetric conditions, in
cylindrical coordinates using a Godunov-type code (see Sec.
\ref{sec:app-code-description}). The grid is of high resolution
with compression towards the disc midplane and towards the
$z-$axis with the goal of having a larger number of the grid-points
in the disc and near the star (see Fig. \ref{grids-3}). After
compression, a typical grid has $N_r=270$ cells in the radial direction and
$N_z=432$ cells in the axial direction. The number of grids
covering the disc in the vertical direction (in the middle of the
disc) is about 200.
 The grid also has
sufficiently high resolution in the $r-$direction. The simulation
region is stretched in the $z-$direction to allow for matter flow in the
corona.

\begin{figure*}
\begin{center}
\includegraphics[width=16cm]{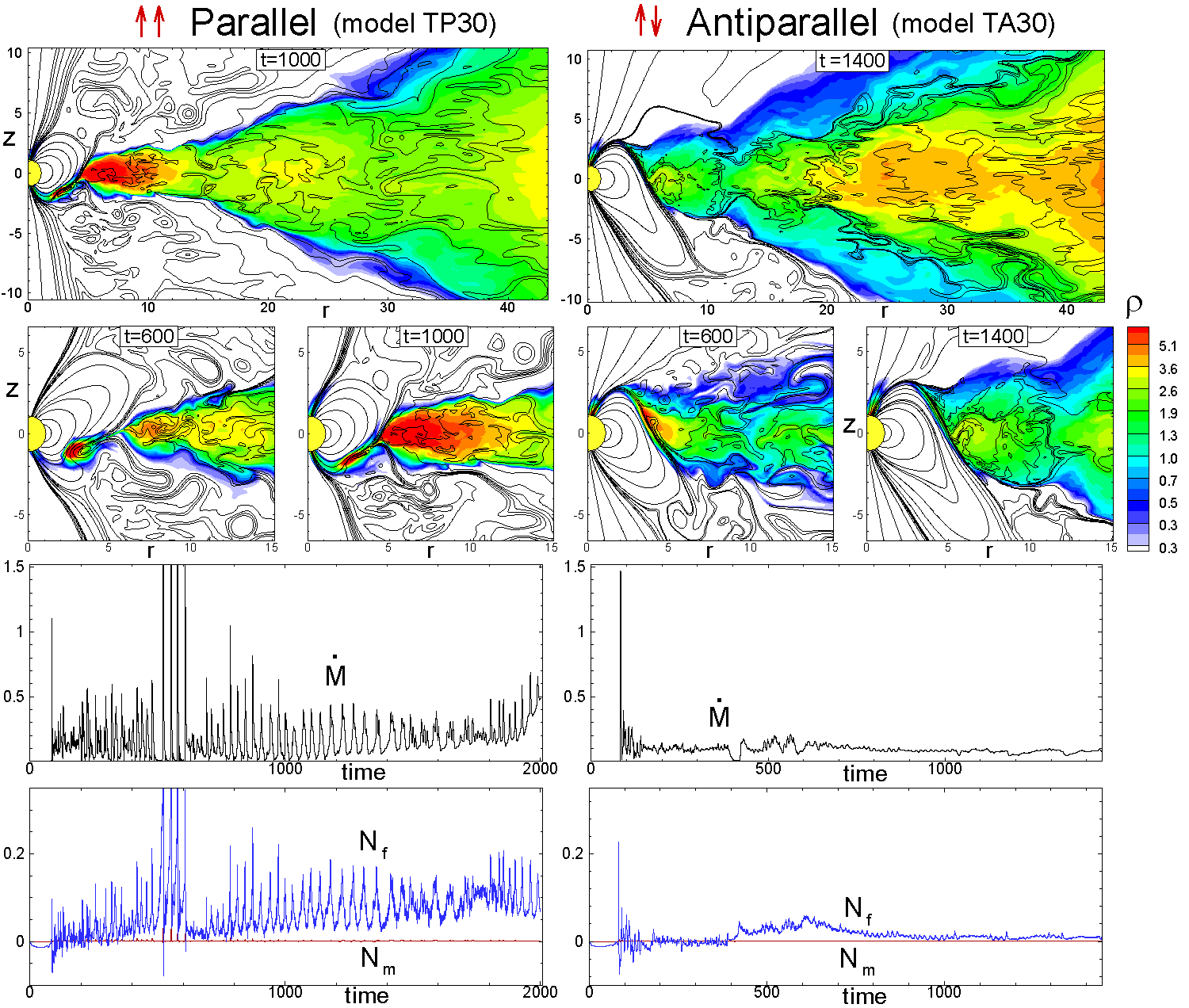}
\caption{\label{accr-mu30} \textit{Left panels:} MRI-driven
accretion at different moments of time for model TP30 where the
seed magnetic field in the disc is parallel to that of the star at
the disc-magnetosphere boundary. The background shows the density,
lines show the poloidal field lines.  \textit{Right panels:} Same
but for the case of antiparallel fields. The top row shows the
flow in the whole simulation region at two moments of time $t=600$
and $1000$. The next row shows matter flow near the star for the
same cases. The 3rd and 4th rows show the temporal evolution of
matter flux onto the star $\dot M$ and torque on the star
associated with matter, $N_m$, and the magnetic field, $N_f$ for
corresponding models.}
\end{center}
\end{figure*}

\subsection{The parameters in different simulation runs}

We ran various simulations given \textit{different initial
configurations} of the field within the disc (see Fig. \ref{init-5}
and Tab. \ref{tab:models}). We observed that similar MRI-driven
turbulence develops in all cases.
 We chose the tapered
field configuration as a base, because it is the simplest after
the vertical field configuration, but most of its magnetic flux is
confined within the disc.

The strength of the MRI-driven turbulence and the subsequent
accretion rate depends on the \textit{value} of the seed poloidal
field in the disc, $B_d$. We observed that if the field is strong
$B_d\gtrsim 0.01$ (see Sec. \ref{sec:app-refunits} for reference
units), the accretion rate is very high, and also the long
wavelength  modes dominate. If the field is weak, $B_d\lesssim
0.0005$) then the accretion rate is very low. We chose the
intermediate value $B_z=\pm0.002$ for almost all simulation runs.
We also performed test runs at higher ($B_d=-0.005$) and lower
($B_d=0.001$) fields. The result also depends on the
\textit{orientation} of the seed field in the disc. Note that the
field of the star at the disc-magnetosphere boundary is positive,
and hence the plus sign for $B_d$ means that the fields have the
same direction.

Our initial setup for the disc and corona \textit{differs} from
the commonly used setup for accretion to black holes, where the
matter is initially confined inside a thick torus of constant
specific angular momentum. The torus gradually evolves into the
disc-type structure (e.g., \citealt{hawl00}). In contrast, our
disc rotates with Keplerian velocity and has a quasi-equilibrium
density distribution from the beginning of simulations (see Sec.
\ref{sec:app-initial-cond}). In addition, matter with a seed field
can flow inward from the side boundary, and this allows for longer
simulation runs (see Sec. \ref{sec:bound-conditions}).

We also performed simulations for different sizes of the
magnetospheres which are determined by the dimensionless parameter
$\tilde\mu$ (see Tab. \ref{tab:models} and Sec.
\ref{sec:dif-moments}).

\subsection{Initial perturbations in the disc}
\label{sec:initial-pert-mri}

 Here, we use our dimensionless
variables (see Sec. \ref{sec:app-refunits}) to  estimate the
number of MRI waves per thickness of the disc. The wavelength of
most unstable, MRI-driven modes is given by $\lambda_{\rm MRI}
\approx 2 \pi {v_{A,z}}/{\Omega_K}$. The full thickness of the
initially isothermal disc is $2 h \approx {2 c_s}/{\Omega_K}$,
where $c_s = \sqrt{p/\rho}$ is the isothermal sound speed in the
disc. Hence, the number of waves per thickness of the disc is
$N_{\rm MRI} = {2 h}/{\lambda_{\rm MRI}} \approx {c_s}/{\pi
v_{A,z}}$. In our simulations (in dimensionless units, see Sec.
\ref{sec:app-refunits}) $c_s = 0.032$, the density in the disc is
$\rho_d \approx 1$,  the Alfv\'en velocity (based on the
$B_d=0.002$ field) is  $v_{A,z} = {B_d}/{\sqrt{4 \pi \rho_d}}
\approx 5.7\times 10^{-4}$. Substituting these values into the
initial formula, we obtain the number of waves per disc and also
the plasma parameter $\beta$ :
\begin{equation}
N_{\rm MRI} = \frac{2 h}{\lambda_{\rm MRI}} = 17
\bigg(\frac{0.002}{B_d}\bigg) , \label{eq:N-mri}
\end{equation}
\begin{equation}
\beta \equiv {2 c_s^2 \over v_A^2} = 6.2\times 10^{3}
\bigg(\frac{0.002}{B_d}\bigg)^2 ~. \label{eq:beta-theor}
\end{equation}
This value of $\beta$ corresponds approximately to the initial
value of $\beta$ in the middle of the disc in our simulations.
Below, we show the development of the MRI-driven turbulence for
one of our models.

\subsection{Development of MRI-driven turbulence in simulations}
\label{sec:initial-development-mri}

Here, we take one of our models (TP30) and show the development of
the MRI-driven flow. We take the  case of the largest
magnetosphere (in our set)
 to investigate whether the large magnetosphere can
hinder the development of turbulence or subsequent inward
accretion towards the star.

When we imposed small random fluctuations of the angular velocity
in the disc, the fluctuations grew. Fig. \ref{beta-rho} (top
panels) shows the distribution of density $\rho$ and $\beta$ at an
early time  ($t=1$). One can see that the density increases
towards larger radii. In our initial conditions (see Sec.
\ref{sec:app-initial-cond}), the density may either increase or
decrease towards larger radii, and it is regulated by a
coefficient in the initial, almost Keplerian, angular velocity
distribution: $\Omega(r)=(1\pm0.02)\Omega_K$, where the `$-$' sign
means
 the density increases outward. We chose this case for all our
simulation runs because it corresponds to softer start-up
conditions and the simulations last longer. Note that the initial
distribution of the plasma parameter $\beta$ is not homogeneous:
it varies between $\beta=20$ at the inner edge of the disc and
$\beta=2\times10^4$ at the outer edge. In the middle of the disc,
at $r=22$, we have $\beta=3\times10^3$ (which roughly
corresponds to the value derived in the previous section). This strong
initial variation in $\beta$ is associated with two factors: (1)
the density increases towards larger radii; and (2) the initial
magnetic field in the disc is a superposition of the strong dipole
field that dominates at the inner regions of the disc, and the
almost homogeneous seed initial field in the disc. The
superposition of fields gives the field $B=0.01$ at the inner edge
of the disc, $B=0.005$ in the middle (at $r=22$) and $B=0.002$ at
the outer edge. Note that in the `torus' initial conditions (e.g.,
\citealt{hawl00}) the $\beta$ also varies across the torus, but not as
much.

Later, we observed that the growth rate of the instabilities is on
the time scale of Keplerian rotations. Fig. \ref{beta-rho} (second
row) shows the density and $\beta$ distributions at time $t=100$,
which corresponds to one Keplerian rotation at $r=21$ (in the
middle of the simulation region). One can see that channel modes
developed at this and smaller radii. The number of modes per
thickness of the disc expected from theory (see Eq.
\ref{eq:N-mri}) is $N_{\rm MRI}\approx 5$ for the inner disc,
$N_{\rm MRI}\approx 8-9$ for the middle of the disc  and $N_{\rm
MRI}=17$ for the outer disc. One can see that at this time, the
instability already developed at the inner disc with few modes
visible in the plot, and the correct number of modes have
developed in the middle of the disc. The external regions show
stretching of the initial perturbations.

The next row in Fig. \ref{beta-rho} shows a later time, $t=300$, which
corresponds to one Keplerian rotation at the outer edge of the
disc $r\approx 44$. One can see that the channel modes have
developed in the whole disc including the outer boundary. Again,
the number of modes roughly corresponds to the predicted
number $N_{\rm MRI}\approx 17$ for this value of the field. Note
that recently developed channel modes in the outer part of the
disc are very ordered, while those in the inner and middle parts
of the disc start to be mixed.

At later times,  $t>500$, the channel modes mix as a result of
interacting with one another, as well as with turbulent matter
that is not in channel streams. Recently, \citet{latt09} suggested
that the `turbulent mixing' is probably the main process which
destroys the channel modes (and not the parasitic instabilities as
suggested earlier, e.g. \citealt{good94}). The bottom  panels
corresponding to $t=1000$ and $t=2000$ show that the flow in most
of the disc is very well mixed and is turbulent. The isolated
channel streams appear in different parts of the disc and
propagate inward, but they do not determine the character of
matter flow at the inner parts of the disc. Formation of isolated
channels which transfer energy to the turbulent matter has been
discussed by \citet{latt09}. Most importantly, the turbulent flow
dominates in the inner parts of the disc which are important for
investigating the disc-magnetosphere interaction.

The left panels of Fig. \ref{beta-rho} show that the density is
gradually redistributed in such a way that the highest density is
in the inner parts of the disc, which corresponds to a
quasi-stationary disc equilibrium.


\section{MRI-driven accretion in cases of parallel and
antiparallel fields} \label{sec:par-antipar}

Our simulations show that the  disc-magnetosphere interaction
depends on the \textit{orientation} of the seed poloidal field in
the disc. If at the disc-magnetosphere boundary the
$B_z-$component of the disc field points in the same direction as
the field of the star, then the field is enhanced at the boundary
initially. Subsequently, the disc matter continues to bring the
field of the same polarity. In the case of antiparallel fields,
the field at the disc-magnetosphere boundary is weaker initially,
and subsequently, disc matter  brings in a field of the opposite
polarity, and this influences matter flow in the inner disc and at
the boundary. Below, we consider two specific models with parallel
(TP30) and antiparallel (TA30) fields and discuss the results in
detail.

\begin{figure}
\begin{center}
\includegraphics[width=8.5cm]{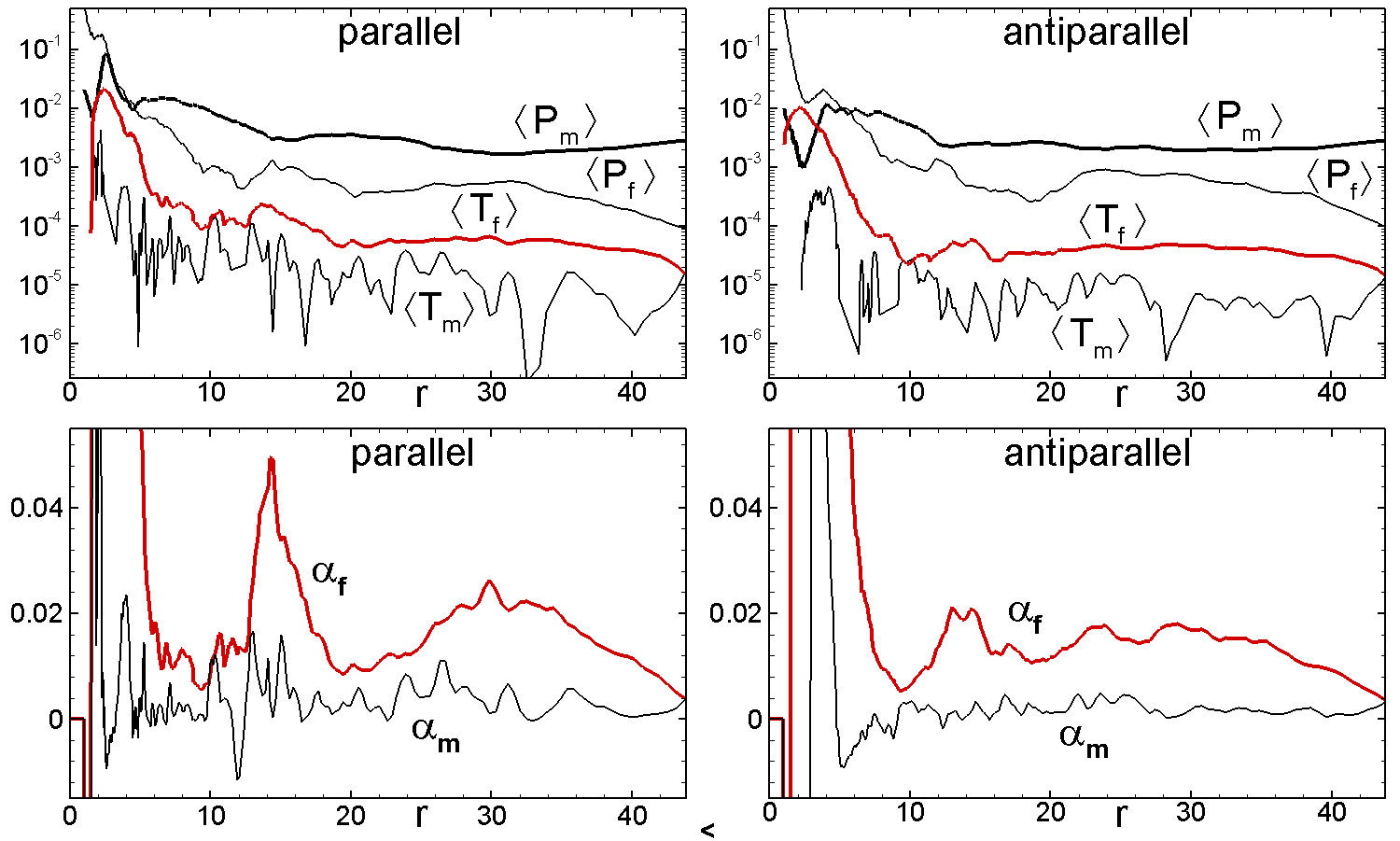}
\caption{\textit{Top panels:} Radial distribution of  matter and
magnetic pressure, $\langle{P_m}\rangle, \langle{P_f}\rangle$, and
stresses, $\langle{T_m}\rangle, \langle{T_f}\rangle$ which were
integrated across the disc, at time $t=1000$. \textit{Bottom
panels:} Same for $\alpha-$parameters associated with matter and
magnetic stresses, $\alpha_m, \alpha_f$. Left and right panels
correspond to models TP30 (parallel fields) and TA30 (antiparallel
fields), respectively.} \label{stress-r-4}
\end{center}
\end{figure}

\begin{figure}
\begin{center}
\includegraphics[width=8.5cm]{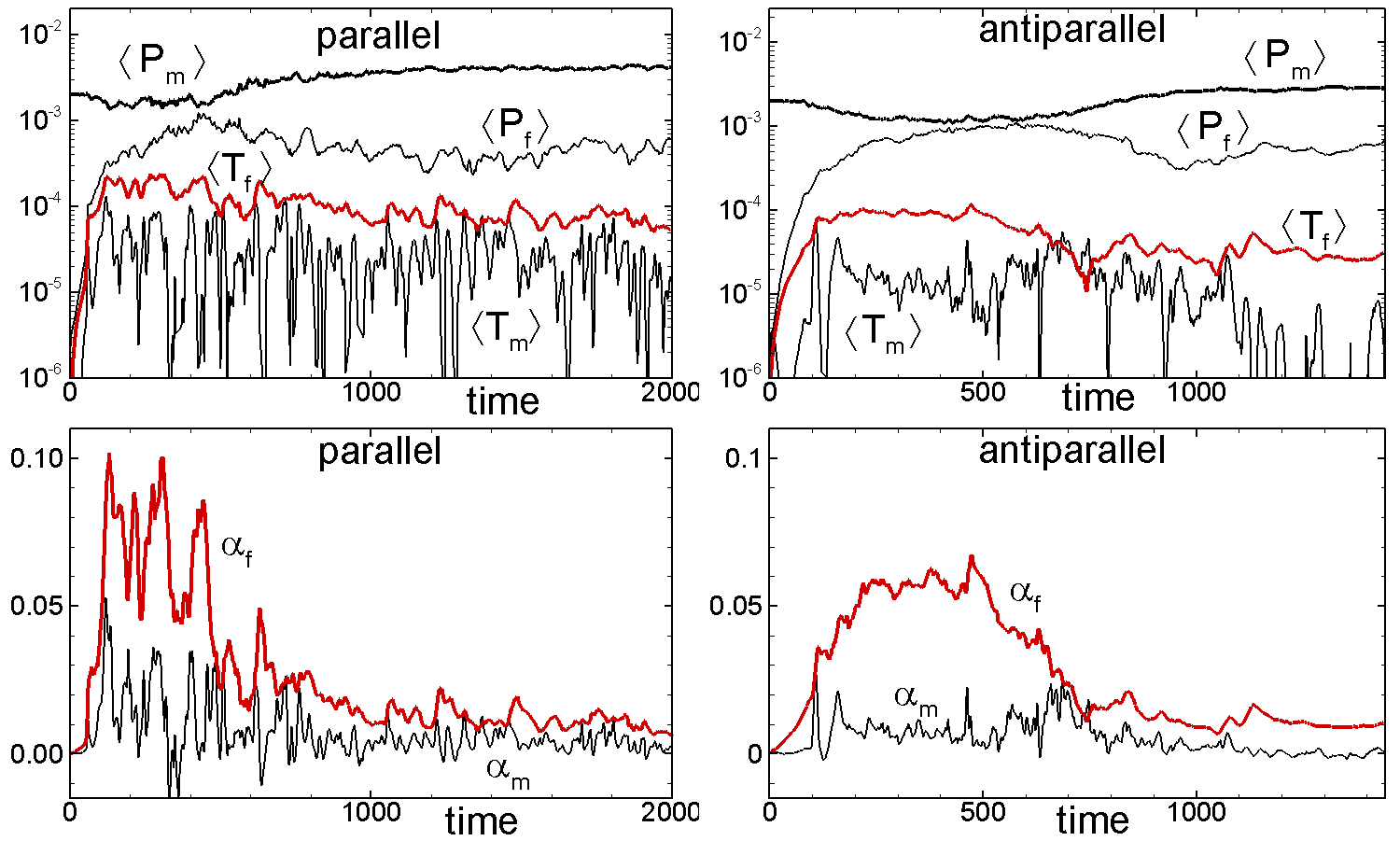}
\caption{\textit{Top panels:} Temporal evolution of  matter and
magnetic pressure, $\langle{P_m}\rangle, \langle{P_f}\rangle$, and
stresses , $\langle{T_m}\rangle, \langle{T_f}\rangle$ taken at the
radius $r=20$ and integrated across the disc. \textit{Bottom
panels:} Temporal variation of $\alpha-$parameters associated with
matter and magnetic stresses, $\alpha_m, \alpha_f$. Left and right
panels correspond to models TP30 (parallel fields) and TA30
(antiparallel fields).} \label{stress-t-4.png}
\end{center}
\end{figure}

\begin{figure}
\begin{center}
\includegraphics[width=7.0cm]{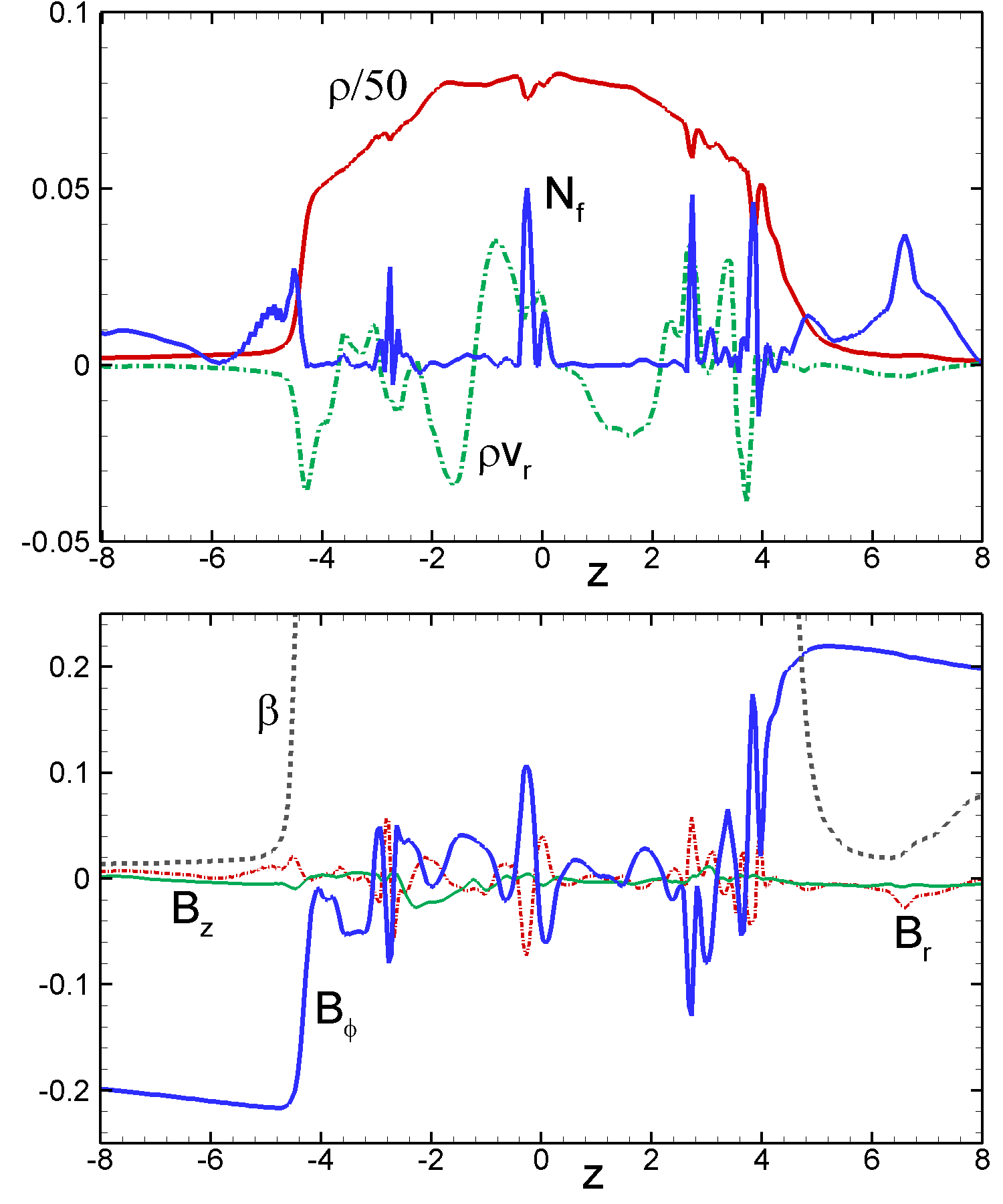}
\caption{\textit{Top panel:} Distribution of radial velocity
$v_r$, specific matter flux $\rho v_r$, specific magnetic torque
$N_f$  and the density $\rho$ across the disc at radius $r=20$ and
time $t=1000$ in model TP30. The density is decreased by a factor
of 50 for scaling, and $N_f$ is increased by a factor of 4.
\textit{Bottom panel:} same as above, but for three components of
the magnetic field and plasma parameter $\beta=8\pi P_m/B^2$. The
plasma parameter varies between small values outside the disc, up
to $\beta>>1$ inside the disc.} \label{z-distr-2}
\end{center}
\end{figure}

\subsection{Matter flow, matter flux and torque}
\label{sec:matter-flow-fluxes}

We observed that in both models, MRI-driven turbulence develops in
the disc and matter flows inward. The top rows of Fig.
\ref{accr-mu30} show accretion on large and small scales at two
times for model TP30 with parallel fields (left side of the plot)
and model TA30 with antiparallel fields (right side). We observed
that in both cases, the disc moves inward due to MRI-driven
turbulence, but it is stopped by the magnetic pressure of the
star. Then matter is lifted above the closed magnetosphere and
accretes to the stellar surface in funnel streams. We observed
that the closed magnetosphere initially expands towards one side,
while matter accretes from the other side. Initial penetration
of the disc matter through the closed magnetosphere is accompanied
by strong reconnection events. Note that the disc is thicker in
the case of antiparallel fields.

We calculated matter fluxes and torque at the surface of the star:
\begin{equation}\label{eq:mdot-star}
\dot M = \int d {\bf S}\cdot {\bf F}_m ~, \quad\quad {\bf
F}_m=\rho{\bf v}_p~,
\end{equation}
\begin{equation}\label{eq:torque-total}
N = \int d {\bf S}\cdot ({\bf N}_{\rm m} + {\bf N}_{\rm f})~,
\end{equation}
where $d {\bf S}$ is the surface area element directed outward;
${\bf N}_{\rm m}$    and  ${\bf N}_{\rm f}$  are torques
associated with matter and the magnetic field:
\begin{eqnarray}\label{eq:torque-m-f}
{\bf N}_{\rm m} &=&  r \rho v_\phi  {\bf v}_p~, \quad\quad {\bf
N}_{\rm f}= - r \frac{B_\phi {\bf B}_p}{4 \pi}~.
\end{eqnarray}

Fig. \ref{accr-mu30} (bottom panels) shows that the matter fluxes
are strikingly different in cases of parallel (left panels) and
antiparallel (right panels) fields. In the case of parallel
fields, matter accretes onto the star in bursts, whereas in the
case of antiparallel fields the accretion is smooth (after the
first burst) and is also at the lower level. We suggest that in
the case of parallel fields, the magnetic flux accumulates at the
disc-magnetosphere boundary and blocks accretion. Matter
accumulates, and later finds a path towards the star.  This leads
to periods of low accretion rate (when accretion is blocked) and
high accretion rate through a burst. For antiparallel fields
fields of opposite polarity frequently reconnect at the
disc-magnetosphere boundary, opening the path for smoother
accretion. It is interesting to note that the `first burst' is
observed in both cases.

We calculated the torque exerted on the star by matter, $N_m$ and
the field $N_f$. We observed that the magnetic torque is much
larger than the material torque and it repeats the pattern of the
matter flux, $\dot M$.   The torque is positive (we changed the
sign compared with the formulae Eq. \ref{eq:torque-m-f} for
convenience of illustration) so the accreting matter spins up the
star. This is what we expect, because the star rotates slowly. The
observed properties of the magnetic and material torques are
similar to those observed in laminar, $\alpha-$discs
\citep{roma02}.

\begin{figure}
\begin{center}
\includegraphics[width=8.5cm]{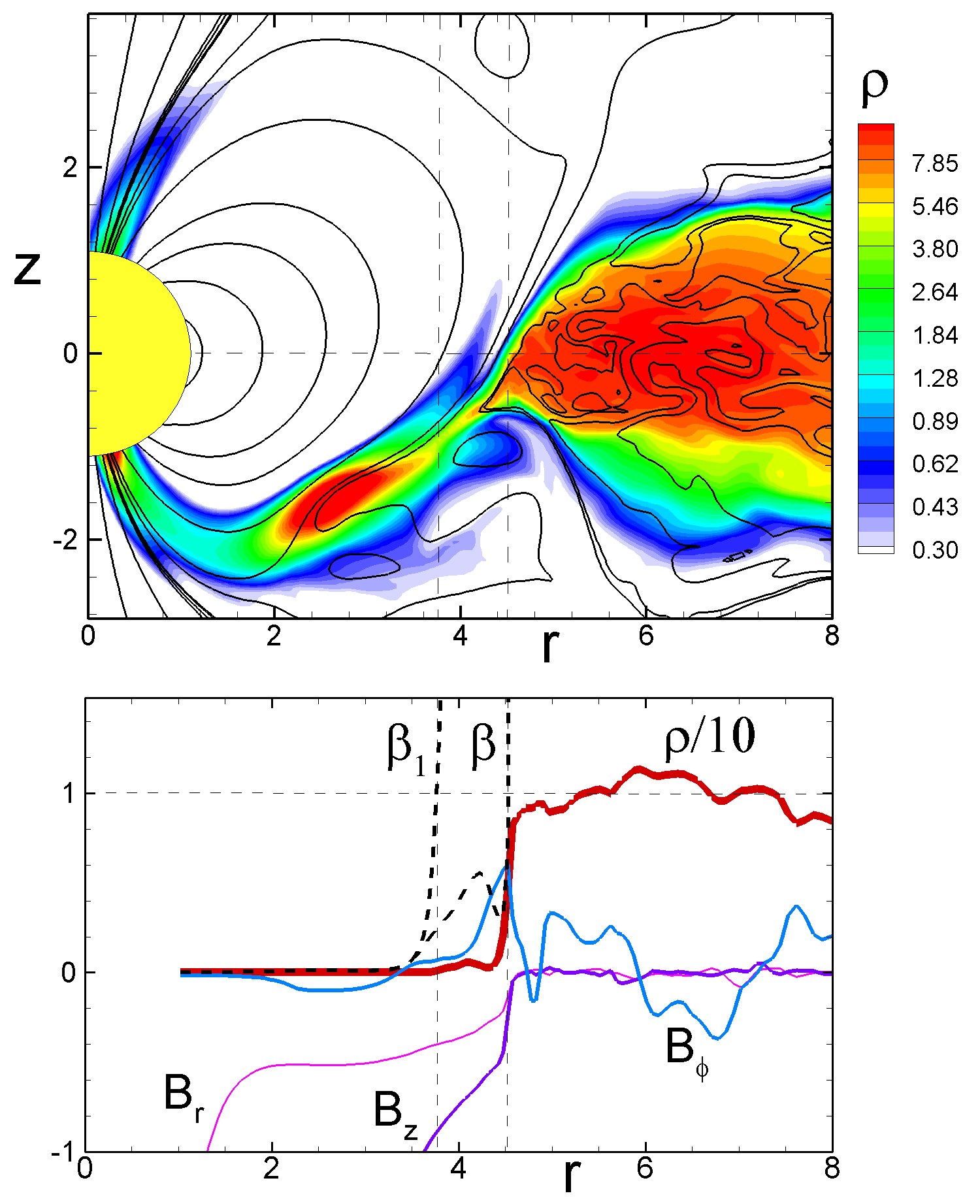}
\caption{{\it Top panel:} close view of the funnel stream at
$T=1000$ in model TP30. {\it Bottom panel:} radial distribution of
density ($\rho$), angular velocity ($\Omega$), plasma parameters
($\beta$, $\beta_1$),  azimuthal component of the field, $B_\phi$,
and radial and z-components of the field ($B_r$, $B_z$). The
density is multiplied by a factor of 0.1 for scaling.}
\label{fun-m-2}
\end{center}
\end{figure}

\begin{figure}
\begin{center}
\includegraphics[width=8.5cm]{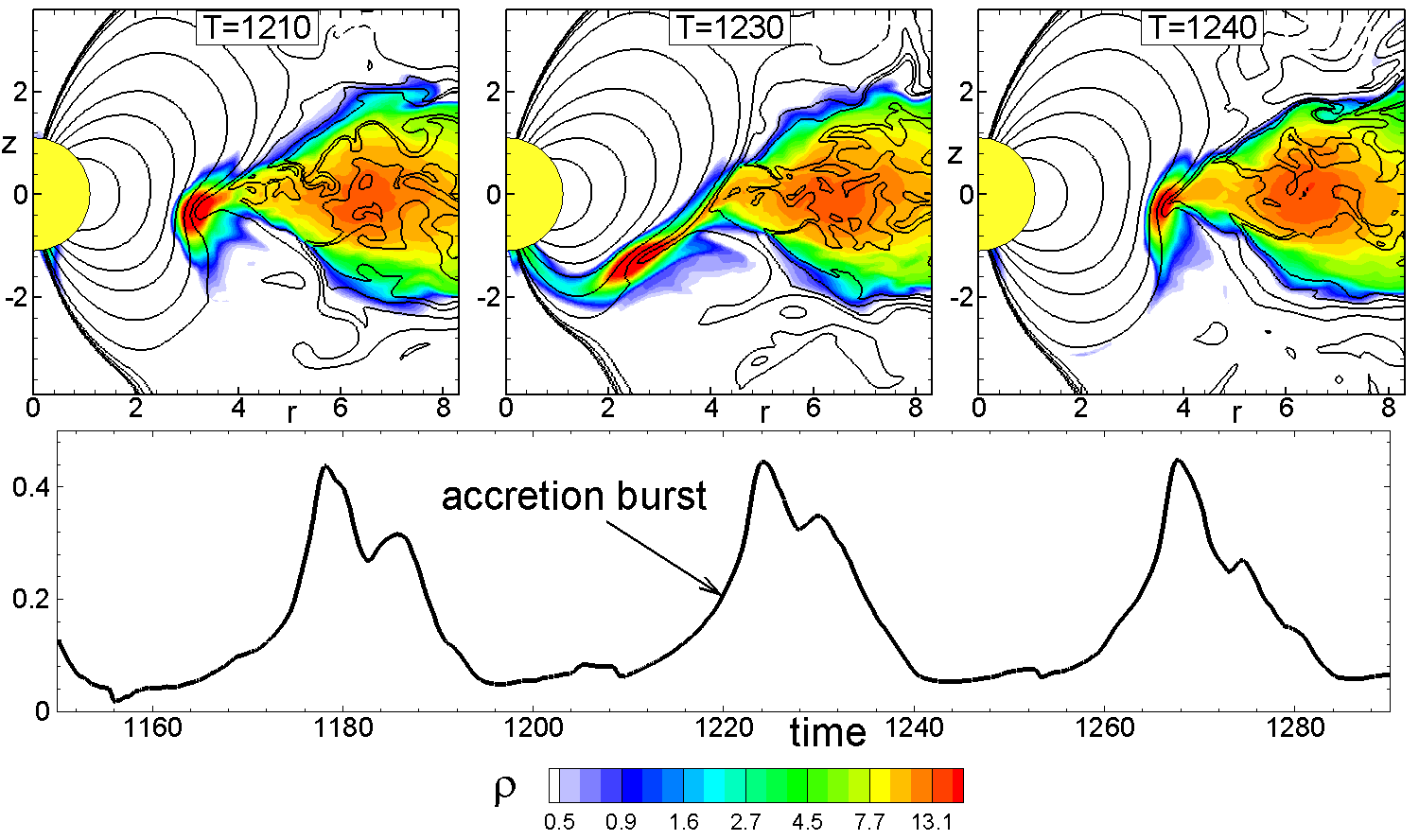}
\caption{{\it Top Panels:} Matter flow before, during, and after
the accretion event  in model TP30, where the accretion rate
enhancement  is determined by the `push' mechanism. The background
shows the density and the lines are selected magnetic field lines.
{\it Bottom Panel:} Matter flux onto the star during this
accretion event, which is marked by an arrow.}\label{push-4}
\end{center}
\end{figure}

\begin{figure}
\begin{center}
\includegraphics[width=8.5cm]{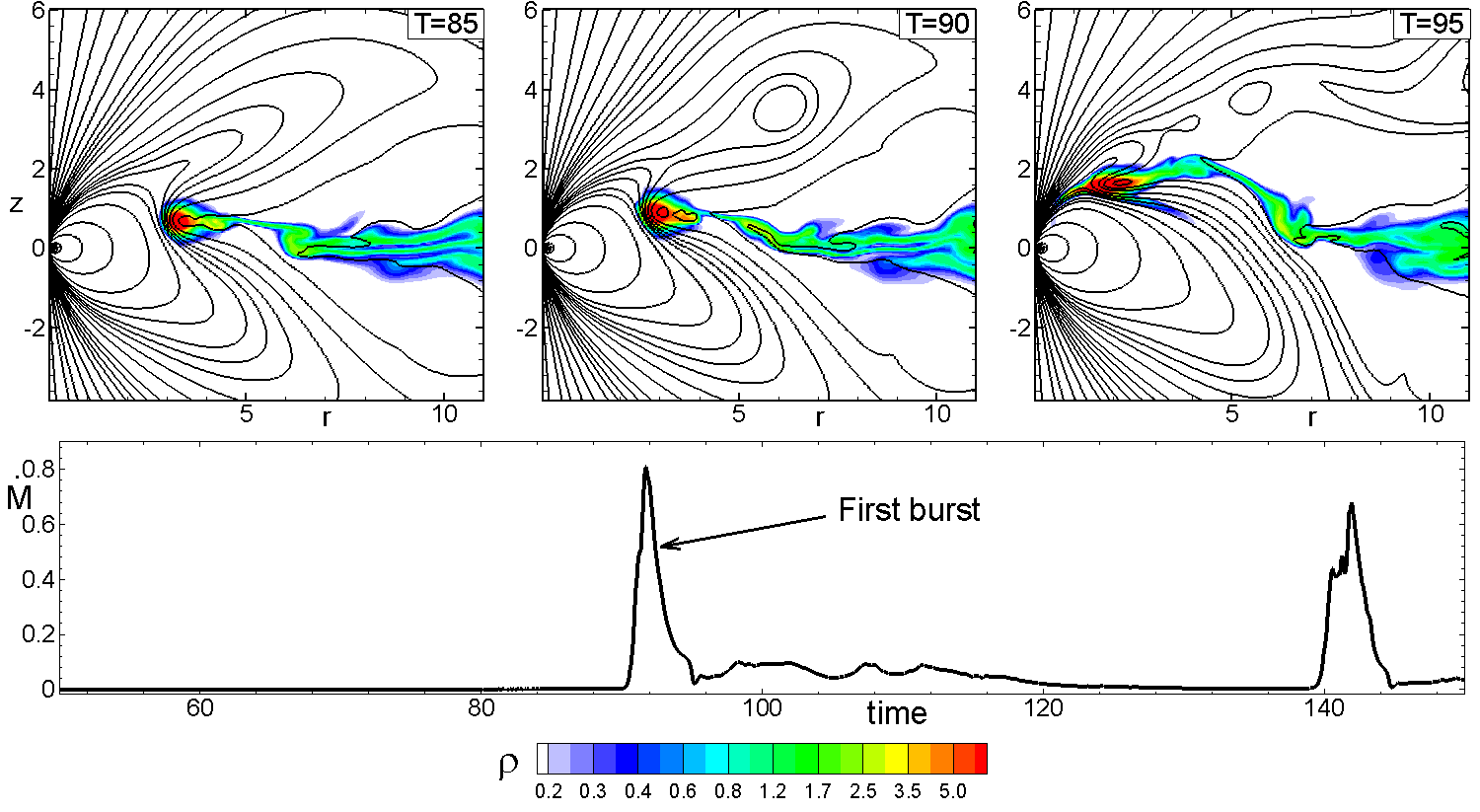}
\caption{\label{recon-4} The figure shows an event of the `first
burst', where the initial stage of the disc
  penetration leads to reconnection of the significant magnetic flux of
  the external magnetosphere. An example is shown for model  TP20.}
\end{center}
\end{figure}

\subsection{Analysis of  stresses and $\alpha-$parameters}
\label{sec:stresses}

Here we analyze the stresses acting inside the disc. We separate
the disc from the low-density corona using one of the density
levels, $\rho_{disc}=0.3$, which is typical for the boundary
between the disc and corona. We integrate matter (subscript `m')
and magnetic (subscript `f') stresses in the $z-$direction and
obtain their distribution as a function of radius in the co-moving
frame:
$$
\langle{T_m}\rangle = \frac{1}{2h} \int dz \rho v_r v_\phi  -
\langle{\rho v_r}\rangle \langle{v_\phi}\rangle
$$
$$
\langle{T_f}\rangle = - \frac{1}{2h} \int dz \frac{B_r
B_\phi}{4\pi}
$$
where
$$
\langle{v_\phi}\rangle = \frac{1}{M} \int dz\rho v_\phi ,
~~\langle{\rho v_r}\rangle = \frac{1}{2h} \int dz \rho v_r ,
$$
are averaged azimuthal velocity and matter flux, and $h=h(r)$ is
the half-thickness of the disc.

The matter and magnetic pressure are
$$
\langle{P_m}\rangle = \frac{1}{2h} \int dz P ,
~~~\langle{P_f}\rangle = \frac{1}{2h} \int dz
\frac{B_{tot}^2}{8\pi} .
$$
The standard $\alpha-$parameters associated with matter and
magnetic stresses are:
$$
\alpha_m =
\frac{2}{3}\frac{\langle{T_m}\rangle}{\langle{P_m}\rangle} , ~~~
\alpha_f =
\frac{2}{3}\frac{\langle{T_f}\rangle}{\langle{P_m}\rangle}.
$$

Fig. \ref{stress-r-4} shows the radial distribution of the
magnetic and matter stresses, pressure, and corresponding
$\alpha-$parameters for models TP30 and TA30. One can see that in
model TP30 (see left panels) the magnetic stress
$\langle{T_f}\rangle$ is about 10 times larger than matter stress
and hence the magnetic stress determines the angular momentum
transport.  The matter pressure dominates over the magnetic
pressure. The magnetic $\alpha-$parameter varies in the range of
$\alpha_f\approx 0.01-0.05$, while the matter $\alpha-$parameter,
$\alpha_m$, is $5-10$ times smaller. Hence, the angular momentum
transport and inward accretion is determined by the magnetic
stress. Note that the size of the stellar magnetosphere
(magnetically-dominated region) is $r_m\approx 4-6 R_*$, and this
region should be excluded from our consideration, because we are
only interested in the stresses in the disc.

The right panels of Fig. \ref{stress-r-4} show corresponding
values for model TA30. In this model the seed poloidal field in
the disc is antiparallel to the stellar field, and initial stress
is somewhat smaller in the inner parts of the disc. We should note
that in this case, the field in the disc is weaker than in the
model TP30, because the oppositely-directed stellar field is
subtracted from the disc field. This leads to smaller magnetic
stress, which is still a few times larger than the matter stress.
The magnetic pressure is smaller than the matter pressure in the
majority of the disc. The magnetic alpha-parameter $\alpha_f$ is
about 1.5 times larger than $\alpha_m$ and the magnetic field is
responsible for the angular momentum transport.

In addition, we investigated variation of stresses and other
variables in time. The $z-$averaged values are taken in the middle
of the disc (at a radius $r=20$). Fig. \ref{stress-t-4.png} (top
panels) shows that the magnetic $\alpha-$parameter ($\alpha_f$) is
larger than the $\alpha_m$-parameter during the whole time, and
hence the magnetic stress determines accretion. Both
$\alpha$-parameters decrease with time.  The value of  $\alpha_m$
has a similar pattern, but at smaller values. The bottom panels of
Fig. \ref{stress-t-4.png} show that all stresses and pressure
values are quite steady with time on average, and only slightly
decrease with time. Note that $\alpha_f$ and $\alpha_m$ decrease
with time, but this is partially because the stresses gradually
decrease with time and partially because the matter pressure
$\langle{P_m}\rangle$ increases.

Results are shown up to $t=2000$ periods of rotation at the inner
boundary ($r=1$), which corresponds to 20 periods of rotation at
$r=22$.

\begin{figure*}
\begin{center}
\includegraphics[width=18.0cm]{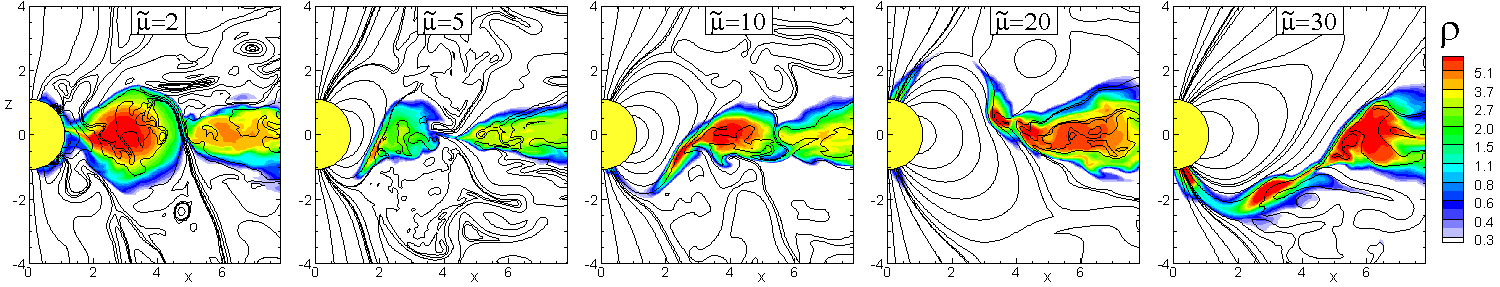}
\caption{Sample plots for accretion onto stars with different
magnetosphere sizes at time $t=700$. Parameter $\tilde\mu$ varies
(from left to right) as $\tilde\mu = 2, 5, 10, 20, 30$.}
\label{dip-diff-5}
\end{center}
\end{figure*}

\section{The disc-magnetosphere interaction} \label{sec:disc-mag}

Here, we investigate in greater detail processes at the
disc-magnetosphere boundary.

\subsection{Where the disc stops}

We take as an example the model TP30 and one of the moments in time
when matter flows along the funnel towards the star.
 Fig. \ref{fun-m-2} (top panel) shows the inner part of the simulation
region. One can see that the disc is stopped at some distance from
the star and matter flows into the funnel stream. The conductivity
is high, and some matter is dragged into the funnel stream.  The
matter energy-density is still higher than the magnetic energy in
the funnel, but the magnetic field imposes some tension force,
which acts in the direction opposite to the flow. This stress is
released due to reconnection inside the funnel stream.
  Such mini-reconnection events are always observed in the
funnel streams, in particular in cases of parallel fields.

The bottom panel of  Fig. \ref{fun-m-2} shows a linear
distribution of different variables in the equatorial plane. One
can see that at the disc-magnetosphere boundary the density $\rho$
drops from high values in the disc to very low values in the
magnetosphere. The $B_\phi$ component of the field dominates
inside the disc, but drops at the boundary. The $B_z$ and $B_r$
components are very small in the disc, but strongly increase in
the region of the magnetically-dominated magnetosphere. Note that
the $B_r-$component is quite large because the magnetosphere is
pushed upward by the funnel stream. Fig. \ref{fun-m-2} also shows
the equatorial distribution of plasma parameter $\beta = 8\pi
P_m/B^2$ and the modified parameter $\beta_1$ based on the total
material stress:

\begin{equation}
\beta_1 = \frac{P_m +\rho v^2}{B^2/8\pi}~ . \label{eq:beta-beta1}
\end{equation}

\noindent One can see that the disc-magnetosphere boundary is
located between the radii where $\beta=1$  and $\beta_1=1$. This
is in accord with earlier simulations of $\alpha-$discs, where the
inner boundary usually coincides with $\beta_1=1$ surface (e.g.,
\citealt{roma02}, \citealt{roma04a}), while the funnel starts at
$\beta=1$ \citep{bess08}.

However, the disc-magnetosphere interaction in the case of
turbulent discs shows many features which are different from those
observed in the case of $\alpha-$discs. One of the differences is
that in the modeling of accretion from $\alpha-$discs the
diffusivity was taken to be of the same order as viscosity, and
both can be small, $\alpha\sim 0.01-0.02$ (like, e.g., in Romanova
et al. 2002, 2003), or larger, $\alpha=0.1$ (e.g.
\citealt{bess08}), and in both cases matter smoothly flows from
the disc to the star.  The MRI-driven turbulence provides
viscosity at the level of $\alpha\sim 0.01-0.1$, but it does
\textit{not} provide the cross-field diffusivity. This is why we
often have the situation that accretion is blocked by the magnetic
flux, and it leads to strong variability. In addition, in cases of
one or another polarity in the disc, the variability is different,
as discussed in Sec. \ref{sec:matter-flow-fluxes} and subsequent
sections.

\subsection{The `push' mechanism of accretion and oscillations}

We often see strong oscillations in the light-curves. Usually,
oscillations are connected with an accumulation of magnetic flux
at the boundary, which generally occurs in cases where the seed
magnetic field has the same polarity as the magnetic field of the star
(cases of parallel fields).

Fig. \ref{push-4} shows one of the episodes of enhanced accretion
observed in model TP30 (the case of parallel fields, see  Fig.
\ref{accr-mu30}). The left panel of Fig. \ref{push-4} shows that
matter accumulates at the disc-magnetosphere boundary. The
magnetic field carried by the inner disc has the same polarity as
the stellar field, and hence they do not reconnect. Matter bends
the magnetosphere in such a way that accretion towards the star
through the funnel stream becomes impossible. However, when
`enough' matter accumulates, it starts pushing the magnetosphere
forward, so that accretion becomes possible (see middle panel).
This happens because the gravity force becomes larger than the
tension force associated with bending. After downloading matter
onto the surface of the star, the magnetosphere expands and
accretion is blocked again by the negative inclination of the
field lines (see right panel). The bottom panel shows the
corresponding burst in the accretion rate. We call this mechanism
the 'push' mechanism. It is responsible for the majority of bursts
seen in the light-curve of model TP30 (see Fig. \ref{accr-mu30},
left panels) and in many other cases shown below.

\subsection{The `First Burst'}
\label{sec:first-flare}

The accretion rate in the disc may vary with time. If accretion rate
is relatively low, then the magnetosphere expands and the disc is
at larger radii. When  accretion rate increases after a period of
low accretion rate, the inner disc moves forward and compresses
the magnetic flux of the star. Finally, it stops at the point
where  material and matter stresses are comparable. However, the
compressed magnetic flux blocks accretion. In MRI simulations the
diffusivity is very low, and the disc matter cannot easily
penetrate through all this flux.

 Initially, our simulations corresponded to such a situation. We observed that
 matter accumulates in the inner disc (see Fig. \ref{recon-4}, left panel), then
 moves towards the star, forcing  all the external
 flux to reconnect (see middle and right panels).
This event leads to a burst in the accretion rate (see bottom
panel). In addition, one can expect strong X-ray
flare associated with reconnection of significant magnetic
flux.

The phenomenon of the first burst corresponds to our initial stage
of evolution. However, such an initial burst is expected in stars with
strongly varying accretion rate. Note that the
`first burst' is expected in cases of both parallel and
antiparallel fields (see Fig. \ref{accr-mu30}).


\begin{figure*}
\begin{center}
\includegraphics[width=14.0cm]{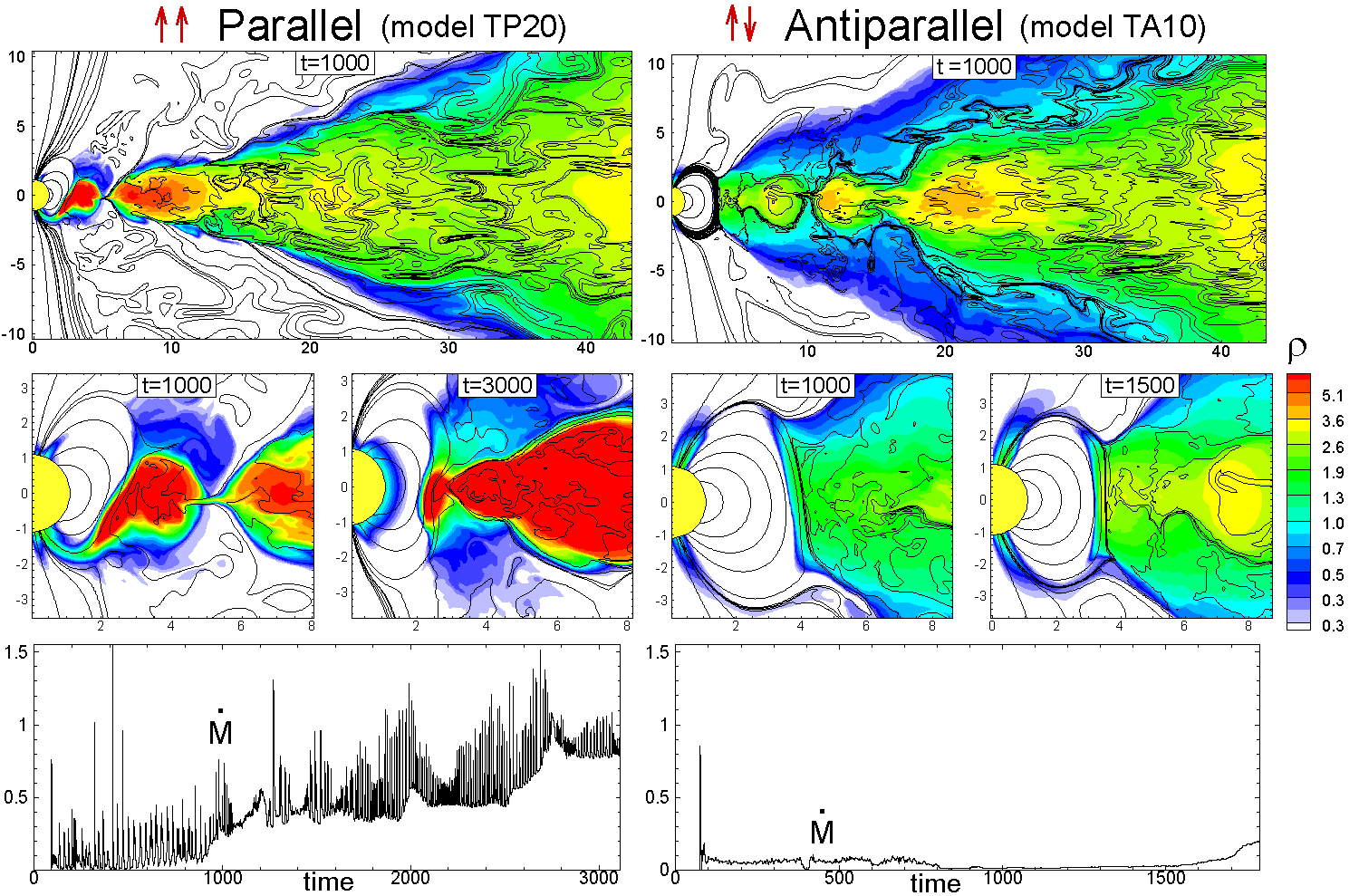}
\caption{Accretion onto a star with a large magnetosphere (models
TP20 and TA20) in cases of parallel (left panels) and antiparallel
(right panels) fields. Bottom panels show accretion rate onto the
star.}\label{accr-mu20}
\end{center}
\end{figure*}

\begin{figure*}
\begin{center}
\includegraphics[width=14.0cm]{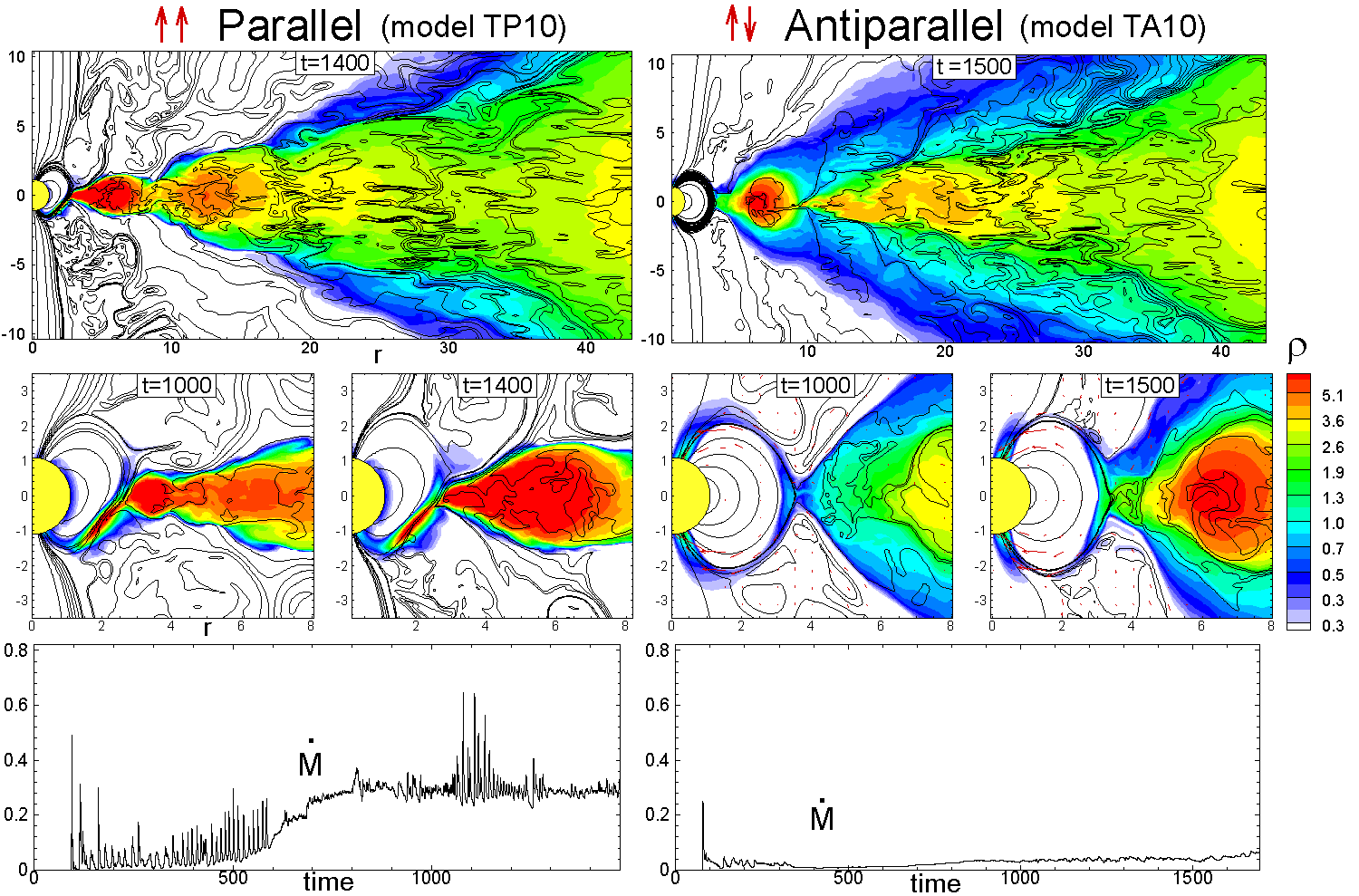}
\caption{Same as in Fig. \ref{accr-mu20} but for a medium-sized
magnetosphere (models TP10 and TA20).}\label{accr-mu10}
\end{center}
\end{figure*}

\begin{figure*}
\centering
\includegraphics[width=14.cm]{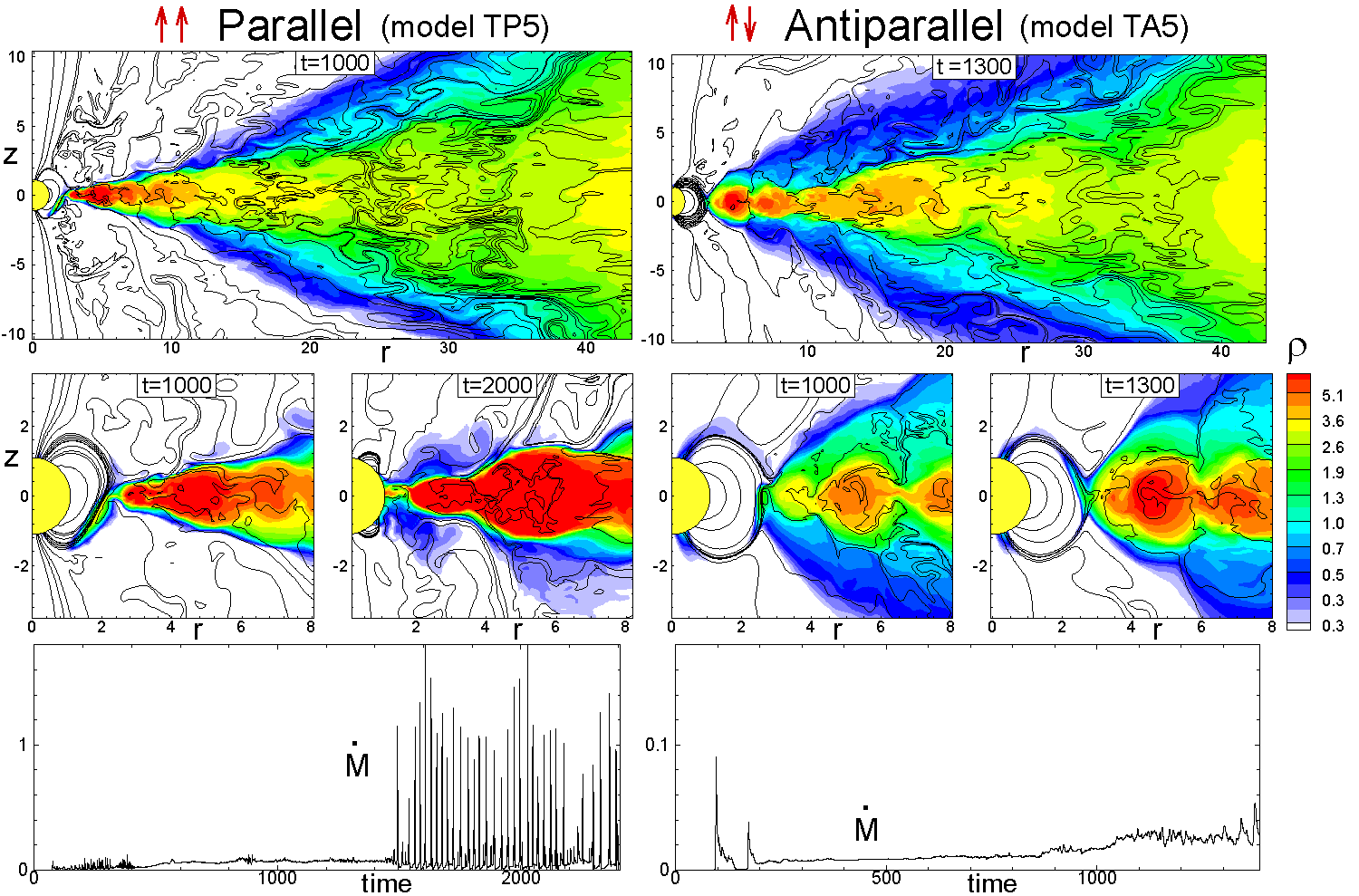}
\caption{Same as in Fig. \ref{accr-mu20} but  for models TP5 and
TA5 where $\tilde\mu=5$. Note the episodic boundary layer
accretion observed in model TP5.}\label{accr-mu5}
\end{figure*}

\begin{figure}
\centering
\includegraphics[width=8.5cm]{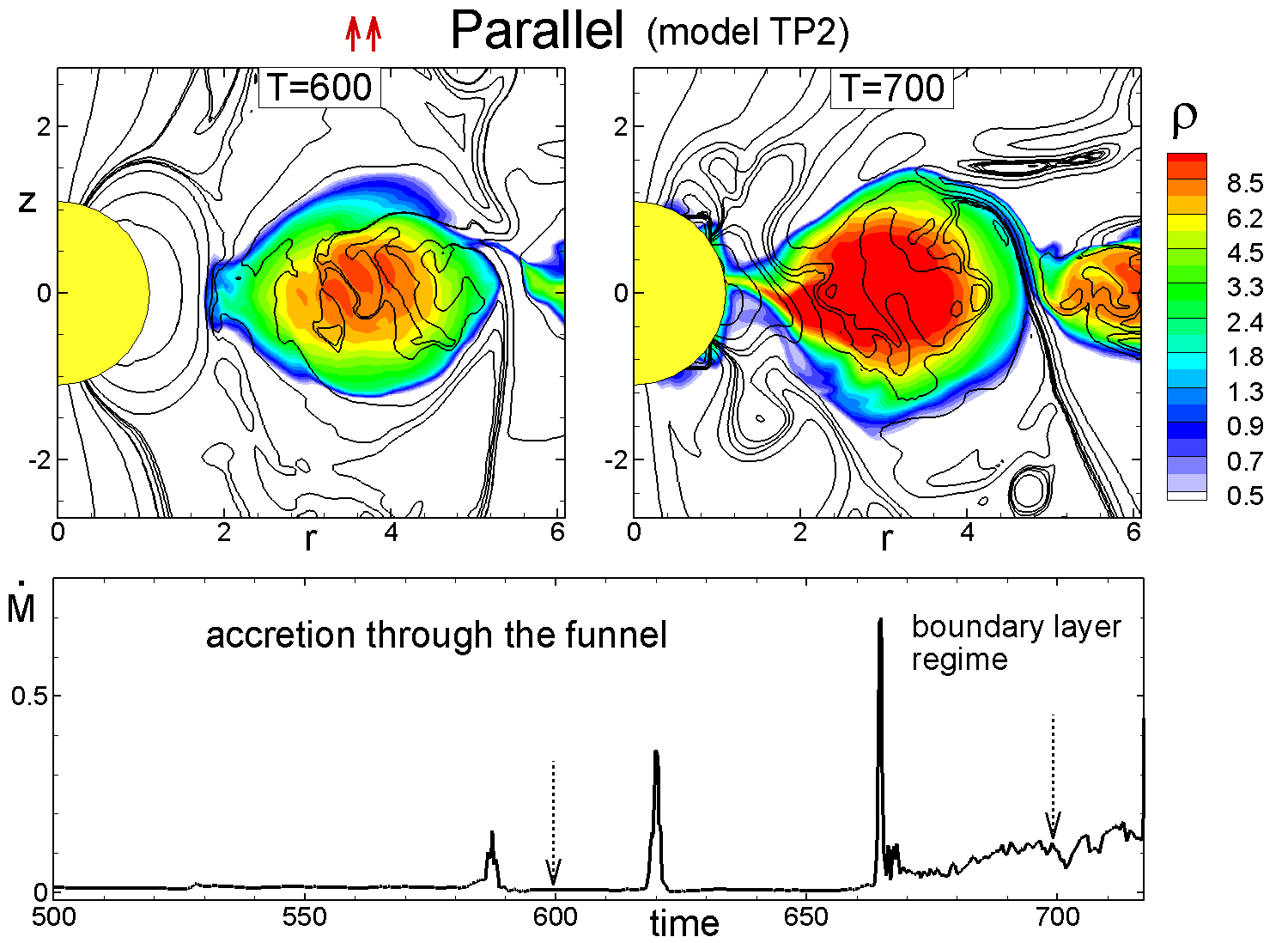}
\caption{\textit{Top panel:} Accretion onto a star with a tiny
magnetosphere $\tilde\mu=2$ and parallel fields (model TP2). Left
and right panels show accretion before and after the onset of the
boundary layer regime. \textit{Bottom panel:} Matter flux onto the
star shows the transition from the funnel flow accretion (at the
left) to the boundary layer regime of accretion (at the
right).}\label{accr-mu2}
\end{figure}

\begin{figure*}
\begin{center}
\includegraphics[width=14.0cm]{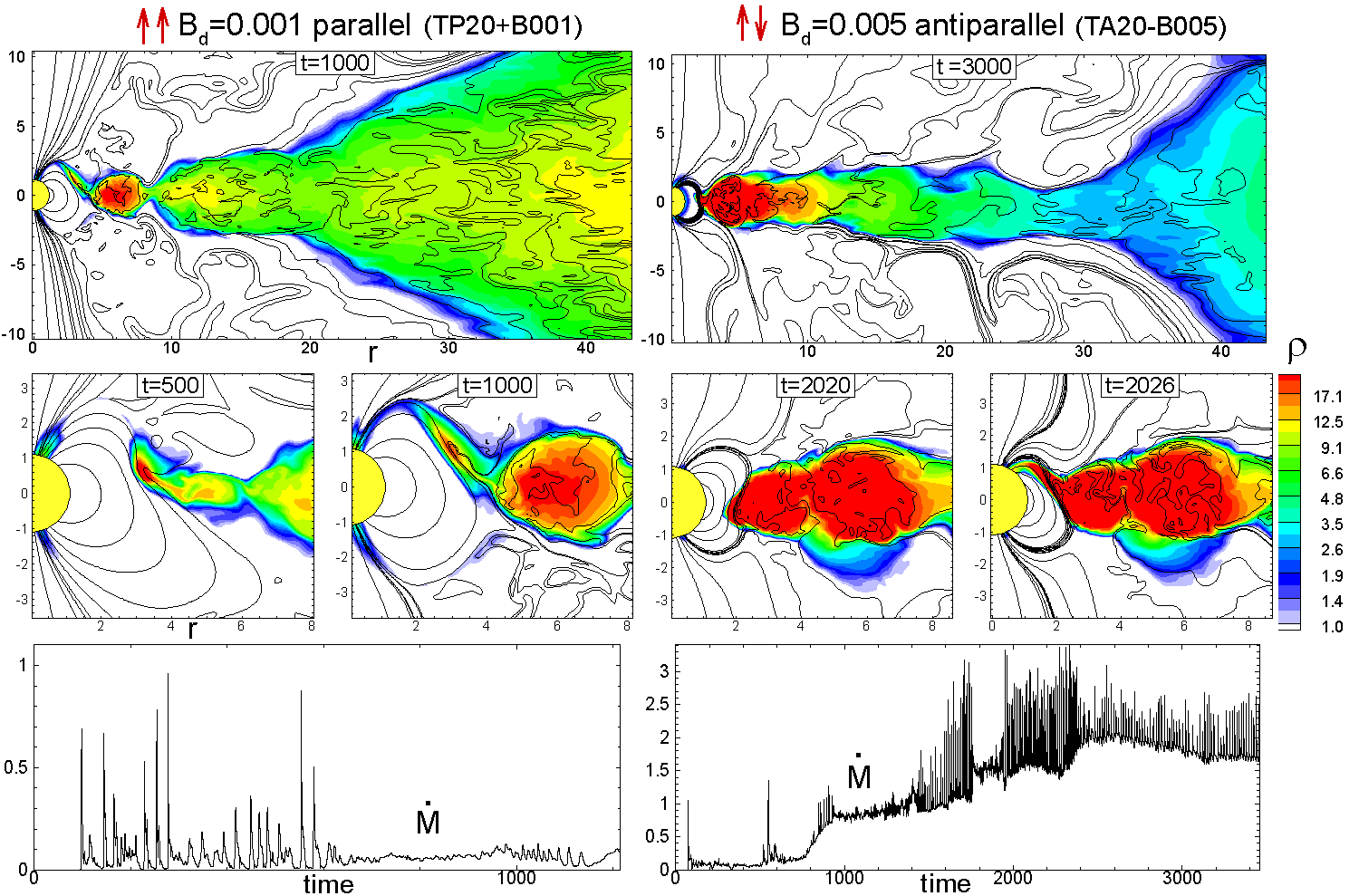}
\caption{{\it Left panels:} Example of accretion and accretion
rate in the case of an intermediate-sized magnetosphere and parallel
magnetic fields (model TP10). \textit{Right panel:} Same, but for
antiparallel magnetic fields in the disc. We chose moments of time
corresponding to bursting and smooth accretion (one per each
case).}\label{accr-mu20-weird}
\end{center}
\end{figure*}

\section{Accretion onto stars with different magnetospheres and accretion rates }
\label{sec:dif-moments}

Above, we described the case of the largest (in our set of models)
magnetosphere, with $\tilde{\mu}=30$. We also calculated accretion
onto stars with smaller-sized magnetospheres at
magnetospheric parameters $\tilde\mu= 20, 10, 5, 2$ (see Tab.
\ref{tab:models}).
 For homogeneity of results
we use the same, tapered topology and the same seed magnetic field in the
disc, which can be parallel (sign `+') or antiparallel (sign `-')
to the field of the star, and has a $z-$component value
$B_d=\pm0.002$.

In two exploratory runs, we fixed the magnetosphere at
$\tilde\mu=20$, but changed the magnetic field in the disc. In one
case, we \textit{increased} the disc field by a factor of 2.5 and used the
antiparallel orientation  ($B_d=0.005$). In the other case, we
\textit{decreased} the field by half and took the parallel orientation
of the field. Below, we analyze results obtained in all these
cases.

Fig. \ref{dip-diff-5} shows magnetospheric flow in cases of the
different sizes of magnetospheres. For homogeneity, we chose the
case of parallel fields, $B_d=0.002$ and show data for the moment
in time $t=700$. One can see that the size of the magnetosphere
decreases with $\tilde\mu$, as expected. The funnel stream forms
from either top or bottom side of the magnetosphere. In the case
of the largest magnetosphere ($\tilde\mu=30$), we observed that
the funnel formed on one side and stayed there (it moved from the
bottom for TP30 and from the top for TA30) during the whole
simulation run. In cases of a smaller magnetosphere, the funnel
flips between the top and bottom sides, although this happens only
a few times per simulation run. The density is usually enhanced at
the inner part of the disc, excluding the case of the smallest
magnetosphere at $\tilde\mu=2$, where the disc accretes directly
through the equatorial plane, that is, in the boundary layer
regime.

In Sec. \ref{sec:par-antipar} we performed an analysis of stresses
for models RP30 and TA30 and observed that the accretion rate
strongly oscillates in the case of parallel fields, but is much
smoother and smaller in the case of antiparallel fields. The
question is, how general is this behavior?  We realize that
initially, the disc is threaded not only with the seed magnetic
field, which we introduce at the level of $B_d=\pm 0.002$, but
also with the magnetic flux of the star, and this flux is larger
in cases of larger magnetospheres and can dominate, or can be
negligibly small, in cases of smaller magnetospheres. Either way,
we expect that the total amount of stress in the disc is different
in cases with different $\tilde\mu$. Hence, with changing
$\tilde\mu$, we also change the total amount of stress in the
disc. Below we investigate the situation and derive general trends
for the value and variability of the accretion rate in different
cases.

\subsection{Matter flow and accretion rates in different models}

First, we show the character of matter flow and accretion rates
obtained in different simulation runs. In figures
\ref{accr-mu20},\ref{accr-mu10}, and \ref{accr-mu5}, we show
results in the same style, where the left and right hand sets of figures
correspond to the cases of parallel and antiparallel fields,
respectively.

Fig. \ref{accr-mu20} shows results of simulations for models TP20
and TA20 ($\tilde\mu=20$). One can see that the MRI-driven
instability has developed in both cases, and in the case of
parallel fields matter accretes through episodes of matter
accumulation and accretion, and the `push' mechanism of accretion
is observed. Simulations show that the accretion rate is high and
strongly oscillating (see left bottom panel of Fig.
\ref{accr-mu20}). In the right panels we see that in the
antiparallel case, the disc is thicker and not as dense. Matter is
accumulated near the star in a thick layer, and accretes onto the
star, but the accretion rate is lower than in the case of parallel
fields, and no oscillations are observed, excluding the first
burst discussed in Sec. \ref{sec:first-flare}

Fig. \ref{accr-mu10} shows a case with a smaller magnetosphere,
$\tilde\mu=10$ (models TP10 and TA10). One can see that in cases
of parallel fields the disc near the star is thin and the density
increases towards the star. Matter accretes through funnel
streams. The matter flux at the star has multiple oscillations up
to $t\lesssim 600$, then it increases and there are fewer
oscillations. We suggest that there are fewer oscillations because
the magnetosphere is smaller and it is easier for matter to climb
the magnetosphere and form a quasi-steady flow. In the case of
antiparallel fields, the density in the disc decreases towards the
disc-magnetosphere boundary. The disc is thicker and the accretion
rate is lower. We analyze the stresses and discuss possible
reasons below.

Fig. \ref{accr-mu5} shows the case of an even smaller
magnetosphere, $\tilde\mu=5$ (models TP5 and TA5). One can see
that in the case of parallel fields, matter smoothly accretes
through the funnel stream, up to time $t\approx 1400$. However,
afterwards the accretion becomes bursty, because matter reaches
the surface of the star episodically and accretes through the
\textit{boundary layer} (see left bottom panel of Fig.
\ref{accr-mu5}). The episodes of accretion are followed by an
expansion of the magnetosphere and an accumulation of matter. Then
the process repeats, giving multiple bursts observed in the
figure. In the case of antiparallel fields (right panels in the
figure), the density in the disc decreases towards the star, and
the geometry is similar to that observed in model TA10. The
accretion rate  is quasi-stationary, like in the TA10 case.

Fig. \ref{accr-mu2} shows an example of a simulation in the case
of the smallest magnetosphere considered in our paper,
$\tilde\mu=2$. The figure shows only the case of parallel fields
(model TP2), where we observed the quasi-steady regime of
accretion through the boundary layer (top right panel). In the
case of antiparallel fields, accretion is similar to that in model
TA5, but the magnetosphere is somewhat smaller. Interestingly,
when the disc accretes through the boundary layer, the magnetic
field of the star is compressed by the disc and obtains a complex
geometry.
\smallskip

There are a few factors which determine the differences in
matter flow and accretion rates seen in different models:

\begin{itemize}

\item Different orientation of the disc field relative to the
stellar field (parallel and antiparallel cases) leads to different
magnetic topologies at the boundary and can influence matter flow
at the disc-magnetosphere boundary;

\item In cases of same/opposite orientations of the field, the
magnetic stress is higher/lower at the boundary;

\item In cases with lager/smaller $\tilde\mu$, the stellar
magnetic field contributes more/less to the total magnetic energy
of the disc, so that the accretion rate is different.

\end{itemize}

Often, these factors act simultaneously, and it is difficult to
separate one from another. To investigate the dependence of
processes on the accretion rate, we chose the magnetosphere with
$\tilde\mu=20$, and  performed two additional runs. In one of
them, we increased the field in the disc by a factor of 2.5:
$B_d=-0.005$. In the case of parallel fields, we expected a higher
accretion rate and bursty accretion. This is why we chose
antiparallel fields, where the situation is less clear (the model
TA20-005). In another test case, we decreased the field by a
factor of two: $B_d=0.001$. We know that in the case of
antiparallel fields we would again observe a very low accretion
rate. Thus, we chose the case of parallel fields, where the lower
accretion rate could yield a different result (model TP20+001).

Fig. \ref{accr-mu20-weird} shows these two cases where the left
and right panels correspond to these test models.  We observed
that for $B_d=0.001$ (left panels), the accretion rate in the disc
is lower compared with the base case (TP20) and the density in the
inner parts of the disc is lower as well. We observed that
accretion is bursty during about 600 rotations, but later a steady
funnel stream is formed. The accretion rate onto a star is a few
times lower compared with the TP20 case. Comparison of these two
cases shows that for the persistent bursty accretion (observed in
TP20), one needs higher accretion rate, at which the magnetosphere
is bent and traps the accreting matter. Hence, the parallel field
configuration is favorable for matter trapping and bursty
accretion, but it is not the only condition.

Fig. \ref{accr-mu20-weird} (right panels) shows that a stronger
magnetic field in the disc generates stronger MRI-driven
turbulence, and the accretion rate is much higher than in any of
the above models. We observed that in such a situation, the inner
disc bends the magnetosphere; the `push' mechanism is switched on
and we observe the persistent bursty accretion. This is not
typical for the case of antiparallel fields in above models. We
conclude that the bursty accretion can appear in the cases of both
parallel and antiparallel fields if the accretion rate is high. We
should note, however, that this is a special case where the
accretion rate is very high, so that the initial rapid accretion
strongly reconstructs the disc.

\begin{figure*}
\centering
\includegraphics[width=14.cm]{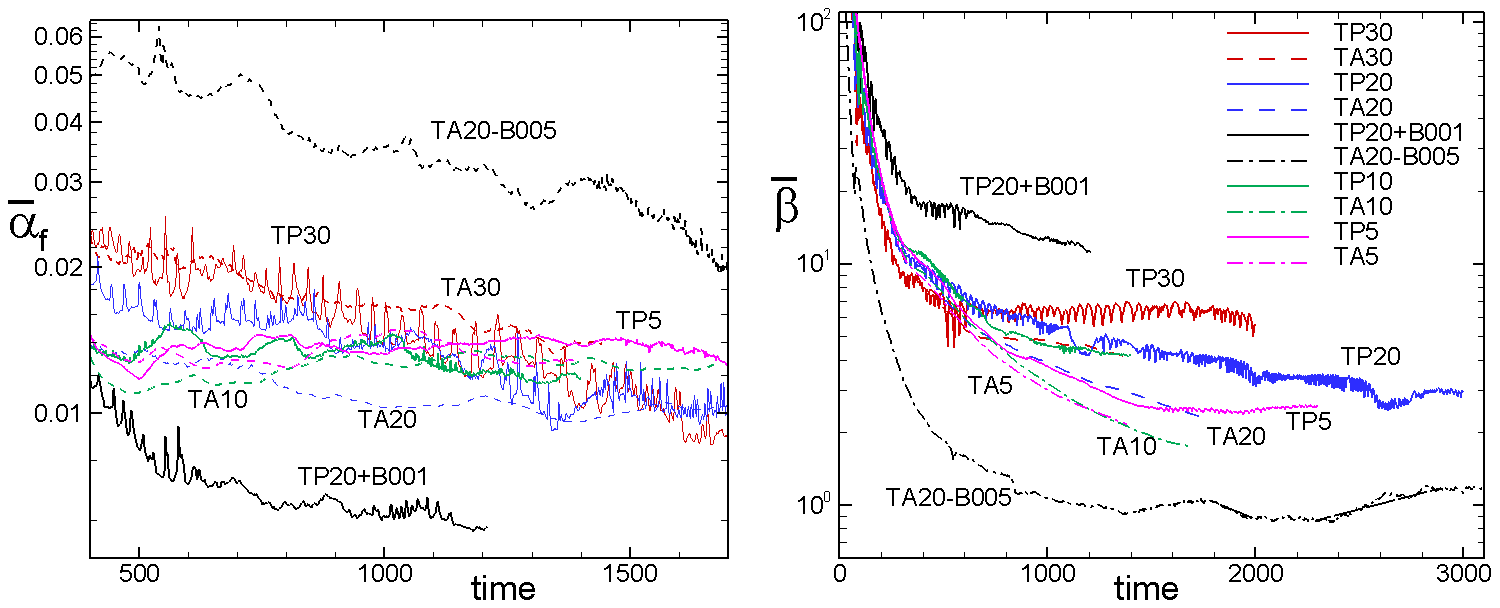}
\caption{\textit{Left Panel:} The magnetic parameter $\alpha_f$ is
averaged along the volume of the disc,  $\bar{\alpha_f}$, and is
shown as a function of time. Only part of the simulation time is shown
for clarity of results. \textit{Right Panel:} Volume averaged
plasma parameter $\bar{\beta}$ as a function of
time.}\label{alpha-beta-all-2}
\end{figure*}

In all above cases, we have different initial magnetic flux in the
disc. To compare the total magnetic energy stored in the disc, we
separated the disc using the density level $\rho=0.3$,
integrated stresses and pressure in the whole disc, and calculated
these values per unit volume. Fig. \ref{alpha-beta-all-2} shows
values $\bar{\alpha_f} =
{2/3}\langle{T_f}\rangle/\langle{P_m}\rangle$ and
$\bar{\beta}=\langle{P_m}\rangle/\langle{P_f}\rangle$.
The right panel of the figure shows
that the value of $\bar{\beta}$ is very high in the
beginning, but drops rapidly because the $B_\phi-$field
strongly increases. So, at $t=1000$, $\bar{\beta}$ varies between 3 and 6 for
our main models, and is about 1 and 12 for test models. These
numbers are reasonable for the MRI accretion discs (e.g.,
\citealt{hawl00}). The volume-averaged parameter $\bar{\alpha}$
(at $t=1000$) varies in the range of $0.01-0.016$ for the main models,
and is smaller ($0.006$) and larger ($0.035$) for test models. We
should note that apart from the model TP30, all other models have
quite similar averaged parameters in the disc, so that the
accretion disc properties are not that different in different
cases. Therefore, we can use different runs for analysis of other
characteristics of the disc-magnetosphere boundary, such as
dependence on reconnection and the local stress depression due to
different orientations of the field. The test cases are different
from the main models, but this was the goal - to check the
difference, and we consider and analyze these cases separately.
  Note that the volume-averaged stresses and $\alpha_f-$parameters are
about twice as small compared with those at the disc-magnetosphere
boundary (see next section).

\begin{figure*}
\begin{center}
\includegraphics[width=10.0cm]{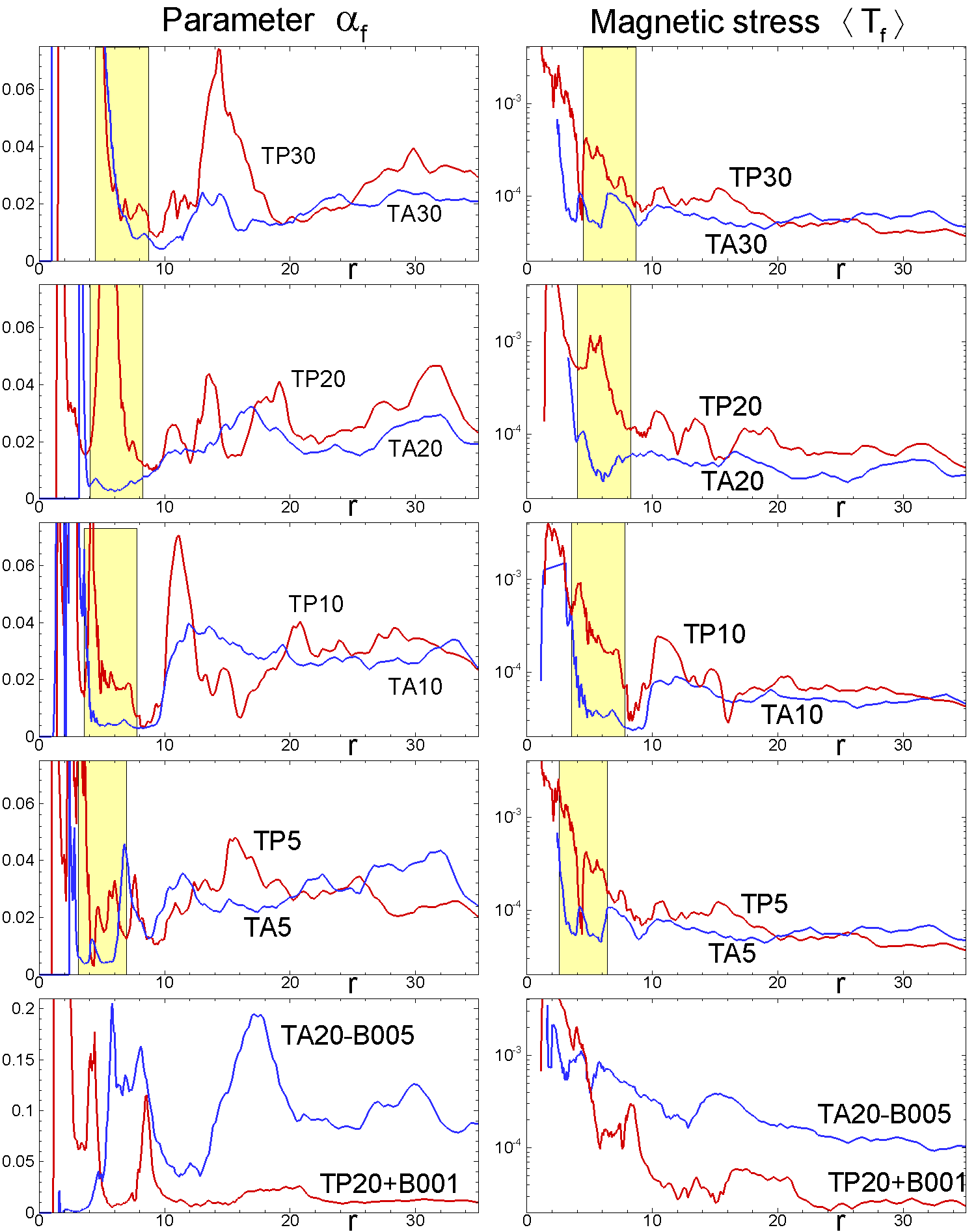}
\caption{{\it Left panels:} Radial distribution of the magnetic
diffusivity parameter $\alpha_f$ in the equatorial plane at sample time
$t=1000$ for models with different magnetospheric sizes
(parameters $\tilde\mu$). {\it Right panels:} same but for
magnetic stresses. The bottom row shows test cases with 2.5 times
larger (model TA20-005) and 2 times smaller (model TP20+001)
values of the magnetic field in the disc, compared with the base
field of $B_d=0.002$.}\label{str-al-r-all}
\end{center}
\end{figure*}

\begin{figure*}
\begin{center}
\includegraphics[width=10.0cm]{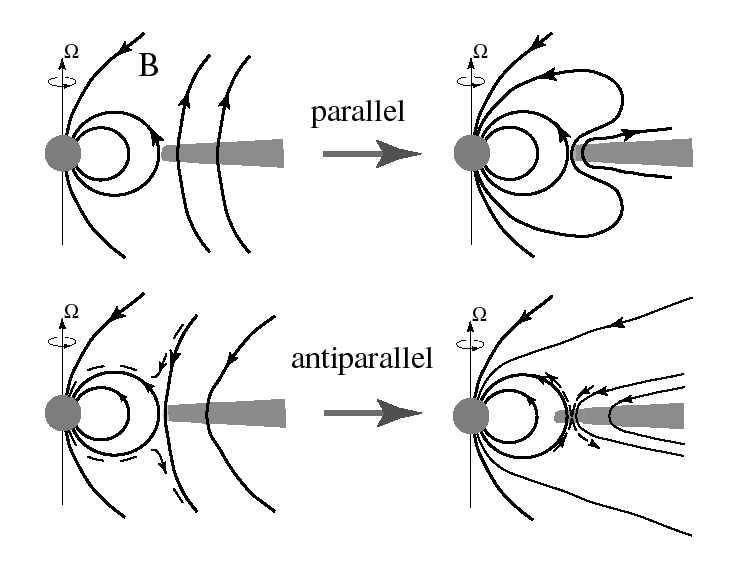}
\caption{{\it Top panels:} Sketch of the disc-magnetosphere
interaction in case of parallel fields (where the disc magnetic
field has the same direction as the stellar field). In this case, the
magnetic flux of the same polarity is accumulated at the boundary, and
the magnetosphere is bent by the disc in the equatorial plane.
{\it Bottom panels:} Same but for antiparallel fields. The dashed
curve in the left panel shows the suggested field topology after
reconnection. The dashed lines in the right panel show the
$X-type$ reconnection point.} \label{sketch-4}
\end{center}
\end{figure*}

\subsection{Analysis of cases of parallel and antiparallel fields}

Here we perform an additional analysis of all cases with the aim
of understanding the role of the field orientation (parallel
versus antiparallel) in different cases.

Sketches in Fig. \ref{sketch-4} show the expected topologies of
the field at the disc-magnetosphere boundary in cases of parallel
and antiparallel  fields. The top panel shows that in the case of
unidirectional (parallel) fields of the disc and the star,
magnetic flux of the same polarity is accumulated at the boundary and
traps the matter. We suggest that this might be a factor which
leads to oscillating accretion. Our simulations of the main models
($B_d=\pm0.002$) show that accretion rate varies from case to
case, but we observe bursty accretion in the case of parallel
fields. The test cases in models TA20-005 and TP20+0.001 show
that the orientation of the field is not the only reason
for bursty accretion. Nevertheless, we clearly see the tendency for
parallel fields to favor bursty accretion.

The right panel in Fig. \ref{sketch-4} shows how the topology of
the field at the magnetosphere boundary changes as the result of
reconnection. It is interesting to note that reconnection between
the field of the star and oppositely oriented field of the disc
leads to the field lines which begin at the star but then stretch
out towards the disc, and often go along the disc. Hence, the
reconnection lets matter flow towards the deeper field lines of
the closed magnetosphere, and therefore acts as an efficient
diffusivity. On the other hand, the distribution of the field
lines after reconnection has a tendency to block formation of the
funnel flow towards the star. We suggest that this factor can be
important in both decreasing the accretion rate and making
accretion smoother. The test model TA20-B005 shows a different
evolution, but this is a somewhat special case. We think that the
opposite polarity of the fields is not a sufficient condition for
smooth accretion. However, at moderate accretion rates, we would
expect less bursty accretion than in the case of parallel fields.

Note that the situation is analogous to the Solar wind interaction
with the Earth's magnetosphere, where one orientation of the solar
wind B-field allows for reconnection and the other one does not
(e.g., \citealt{frey03}).
\smallskip

Another important issue is that the magnetic stress at the
disc-magnetosphere boundary is larger/smaller in cases of
parallel/antiparallel fields. This can lead to a local decrease in
the accretion rate at the boundary. We integrated stresses in the
$z-$direction, and plotted them as a function of radius at
$t=1000$. Fig. \ref{str-al-r-all} (right panels) shows the radial
distribution of magnetic stresses $\angle{T_f}\rangle$, where red
and blue lines indicate parallel and antiparallel fields,
respectively. One can see that at the inner boundary,  stresses
are systematically (about 10 times) larger in cases of parallel
fields. The region where stresses are different is located to the
right of the magnetically-dominated magnetosphere, and has a
typical thickness of $\Delta r_{\rm stress}=4 R_0$. The size of
the magnetosphere decreases with $\tilde\mu$ and hence this region
of different stresses (see yellow-color region in Fig.
\ref{str-al-r-all}) moves slightly to the left. The left panel
shows that the $\alpha-$parameter is also systematically larger in
cases of parallel fields, although the result is not as clean
because matter pressure $\langle{P_m}\rangle$ varies and is in the
denominator of $\alpha_f$. For example, $\langle{P_m}\rangle$ is
higher in cases of parallel fields. Nevertheless, we see
enhancements of $\alpha$ in models with $\tilde\mu=20, 10, 5$. We
conclude that the orientation of the field does change the local
accretion rate in the region of the inner disc that is adjacent to
the magnetosphere.

Fig. \ref{str-al-r-all} also gives a representative number for
$\alpha_f$ in different models. One can see that this parameter
varies in the range of $\alpha_f\approx 0.01-0.04$ with an average
value of $\alpha_f\approx 0.02-0.03$ for all our standard models.
Note that at $r\gtrsim 10$, discs with parallel and antiparallel
fields give similar $\alpha_f$.

The bottom plots of Fig. \ref{str-al-r-all} show our test cases.
One can see that for $B_d=-0.005$, the stress is much larger and
$\alpha_f\approx 0.1$ on average, while for $B_d=0.001$, the
stress is smaller than in our main cases, and $\alpha_f\approx
0.01-0.02$.

\section{Magnetically-dominated corona}
\label{sec:corona}

Initially, we have a low-density corona threaded by the dipole
magnetic field. The differential rotation of the foot-points of
loops (threading the disc and the star) could lead to the
inflation of the field lines and to the expansion  of the corona
\citep{love95}. Both, the coronal expansion and the MRI
instability occur with the same time-scale of Keplerian rotation
at a given radius $r$. However, we observed that the global
inflation of the coronal magnetosphere has been strongly disturbed
by the initial penetration of the inner disc through the external
magnetosphere. We observed that the initial penetration led to a
strong reconnection event (see Fig. \ref{recon-4}) and a large
magnetic island formed and moved outward along the disc. This
island pushed the coronal field lines threading the disc to larger
radii. Subsequent events of reconnection led to the formation of
many more (but smaller) islands, which formed the turbulent
component of the magnetically-dominated corona. The field lines
connecting a star with the external corona inflated and formed an
elongated field structure  propagating up and down. Subsequent
evolution led to the winding of the coronal field lines and to the
formation of a magnetically-dominated corona with a high azimuthal
component of the field. The corona expanded slowly to larger and
larger distances from the disc. We investigated the corona using
one of our typical cases (model TP20). Simulations show that the
magnetically-dominated corona reached the top and bottom
boundaries of the region at $t\approx 3000 P_0$,  which
corresponds to about 30 rotations at the grid center (in the
$r-$direction). The inner parts of the corona, closer to the star,
are turbulent and represent multiple magnetic islands. The
external 2/3 of the disc are threaded with the ordered magnetic
field expanding outward. Interestingly, no disc wind flows from
the disc to the corona along these lines, in spite of the high
inclination of these field lines towards the disc. In addition,
the magnetic field is tightly wrapped in the corona and a high
gradient of magnetic pressure is observed. In spite of that,
neither centrifugally-driven, nor magnetically-driven wind was
formed along these field lines. This issue requires further
investigation.

This magnetic, slowly moving corona resembles the corona formed in
\citet{mill00} simulations. However, in their simulations, the
field in the corona has been generated in the disc and moved
upward as a result of buoyancy. In our simulations, the field
originates from the initial coronal field of the star, but is
strongly wrapped by the rotation of the disc.

\begin{figure*}
\begin{center}
\includegraphics[width=12.0cm]{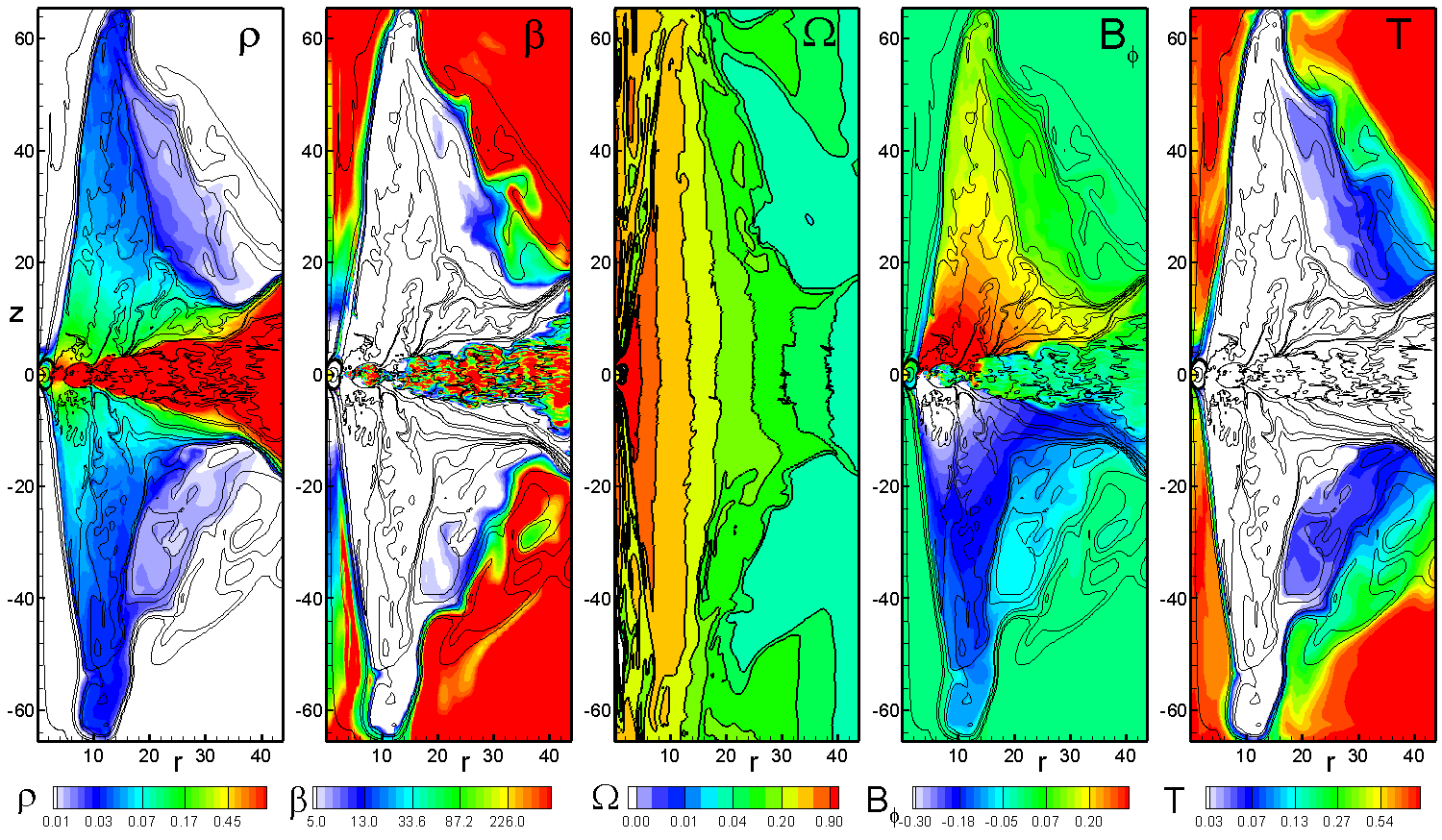}
\caption{Distribution of different values in the whole simulation
region in the model TP20 at time $T=3000$. The lines are sample
magnetic field lines. The color background (from left to right):
density ($\rho$), $\beta-$parameter, angular velocity ($\Omega$),
azimuthal magnetic field ($B_\phi$), and temperature ($T$).}
\label{long-5}
\end{center}
\end{figure*}

 Fig. \ref{long-5} shows different features of the corona.
 The left panel shows that the
corona has low density. Next panel (to the right) shows that the
expanding region of the corona is magnetically-dominated (plasma
parameter $\beta<<1$). The next  plot shows that the corona rotates
approximately with an angular velocity of the disc. Next plot shows
that the azimuthal field $B_\phi$ is strong and that there is an
outwardly directed  gradient of magnetic pressure.  The
azimuthal field  is about 10-100 times stronger than the poloidal
field. The right panel shows that matter in the expanding corona
is cold and hence the corona expands due to magnetic pressure.

 This new type of magnetic structure is not a jet
but instead a magnetic formation filled with the ordered
(mainly azimuthal) and turbulent magnetic field which slowly expands in the
axal direction.

\section{Astrophysical examples}

Table \ref{tab:refval} shows typical reference values. Below are a
few examples where the values obtained in simulations are
converted to dimensional values for typical CTTSs, magnetic CVs and
millisecond pulsars.

\subsection{Application to Classical T Tauri stars}

We consider a few simulation runs relevant to $\tilde\mu=20$ (TP20,
TA20, etc.) and convert dimensionless numbers presented in the
paper to dimensional values.

The reference values shown in Tab. \ref{tab:refval} are relevant
to stars with different $\tilde\mu$, but the magnetic field of the
star $B_\star$ is different for different $\tilde\mu$. For
$\tilde\mu=20$, we obtain an equatorial field:
$B_\star^{eq}=\tilde{\mu} B_0=1000$G (and polar field
$B_\star^{p}=\tilde\mu
* B_0=2000$G). The $z-$component of the seed magnetic field in the
disc is $B_d=0.002 B_0=0.1$G, and the azimuthal component of the
field amplifies up to $B_\phi\approx 0.5-1$G. The magnetic field
of the star at the disc-magnetosphere boundary (say, at
$3R_\star$) is $B=37$G, that is, much larger than the field in the
disc. Fig. \ref{accr-mu20} (bottom panels) shows accretion rate
$\widetilde{\dot M}$ in cases of parallel and antiparallel fields.
In the case of parallel fields we see that the accretion rate
varies in the range $\widetilde{\dot M}\approx 0.1-0.8$ with
bursts up to $1.5$, which corresponds to dimensional values ${\dot
M}_{parallel}=\widetilde{\dot M} M_0 \approx
(0.6-3.6)\times10^{-8} \rm{M}_\odot \rm{yr}^{-1}$ with bursts up
to $8.4\times10^{-8} \rm{M}_\odot \rm{yr}^{-1}$. In model TA20
(with antiparallel fields), the accretion rate is much smaller,
$\widetilde{\dot M}=0.01-0.05$, and the dimensional accretion rate
is ${\dot M}_{antiparallel}=\widetilde{\dot M} M_0 \approx
(0.1-2.8)\times10^{-9} \rm{M}_\odot \rm{yr}^{-1}$. Hence, if the
global magnetic field alternates polarities, then the accretion
rate may vary from higher values and oscillating character to
lower rates with no oscillations. The time-scale between
oscillations is $\Delta t \times P_0=0.37{\rm days}\times 35= 13$
days, and about half of this value for later times. Note that
values of accretion rate will increase if we take a larger
magnetic field of the star and hence larger $B_0$. The dependence
is $\dot M\sim B_\star^2$.

\subsection{Application to millisecond pulsars}

 For $\tilde\mu=20$, we obtain the equatorial field of the star
$B_\star^{eq}=\tilde{\mu} B_0=10^9$G and the polar field
$B_\star^{p}=2\times 10^9$ G. The $z-$component of the seed
magnetic field in the disc, $B_d=0.002 B_0=10^6$G, and the
azimuthal component of the field amplifies up to $B_\phi\approx
10^7$G.   Fig. \ref{accr-mu20} (bottom panels) shows accretion
rate $\widetilde{\dot M}$ in cases of parallel and antiparallel
fields. In the case of parallel fields  the accretion rate varies
in the range $\widetilde{\dot M}\approx 0.1-0.8$, with bursts up
to $1.5$, which corresponds to dimensional values ${\dot
M}_{parallel}=\widetilde{\dot M} M_0 \approx
(0.23-1.8)\times10^{-8} \rm{M}_\odot \rm{yr}^{-1}$, with bursts up
to $3.4\times10^{-8} \rm{M}_\odot \rm{yr}^{-1}$. In model TA20
(with antiparallel fields), the accretion rate is much smaller,
$\widetilde{\dot M}=0.01-0.05$, and the dimensional accretion rate
is ${\dot M}_{antiparallel}=\widetilde{\dot M} M_0 \approx
(0.23-1)\times10^{-9} \rm{M}_\odot \rm{yr}^{-1}$. Hence, if the
global magnetic field alternates polarities, then the accretion
rate may vary from higher values and oscillating character to
lower rates with no oscillations. The time scale between
oscillations is $\Delta t \times P_0=0.46{\rm ms}\times 35= 16$
ms, and about half of this value for later times. Note that the
interval between bursts increases with the spin of the star and
reaches the value of $300$ in the propeller regime \citep{usty11}.
This gives a much longer interval between bursts,  $\Delta
t_{bursts}=138$ ms. In a real astrophysical situation, where the
diffusivity is probably even smaller than in our simulations, one
can expect even longer intervals between bursts. Some millisecond
pulsars show accretion in a flaring regime. For example, the
pulsar SAX J1808.4-3658 shows flaring activity in the tail of the
burst with a quasi-period of 1Hz \citep{patr09}. We suggest that
these flares may be connected with the disc-magnetosphere
interaction in the propeller regime (see also
\citealt{roma04b,roma05,usty06,roma09}).

\section{Conclusions}

We performed axisymmetric simulations of accretion
onto magnetized stars from MRI-driven discs. The main conclusions
are the following:

\begin{itemize}

\item Long-lasting, MRI-driven accretion has been developed in
the disc where the magnetic stresses determine the accretion rate
with equivalent parameter $\alpha_m\approx 0.02-0.04$ (with larger
and smaller values in test models).

\item We observed that the matter-dominated turbulent disc is
stopped by the dipole field of the star at the radius where the
magnetic stress of the magnetosphere matches the matter stress in
the disc. Matter is lifted above the magnetosphere and flows in
funnel stream, which usually picks one of the sides (top or
bottom), and may switch sides a few times during the simulation
run.

\item Processes at the disc-magnetosphere boundary depend on
 the orientation of the poloidal field in
the disc:

~~~\textit{(1).~Parallel fields:} If the field has the same direction
as the stellar field at the boundary, then the
magnetic stress and accretion rate are locally enhanced. However, the field of
same polarity is accumulated at the boundary and
matter is trapped.
 This matter bends the field lines of the
magnetosphere near the equator, and accretion is prohibited.
Later, matter is accumulated and flows to the star through a
temporary funnel stream, and the magnetosphere expands again. Then
the process repeats. This leads to strong oscillations of the
accretion flux on the star. Such oscillations appear as the result
of a relatively high accretion rate, as determined by MRI, and low
diffusivity at the disc-magnetosphere boundary. In a test run with
lower accretion rate ($\alpha_m\approx 0.01$), temporary
oscillations were followed by quasi-stationary accretion through
the funnel flow.

\smallskip

~~~\textit{(2).~ Antiparallel fields:}~If the fields of the star
and the disc have opposite directions, then reconnection of the
field at the disc-magnetosphere boundary can help matter penetrate
towards deeper layers of the closed magnetosphere. At the same
time, the magnetic  flux cancellation lowers the magnetic stress
locally, within a few stellar radii of the magnetosphere. This
leads to lower accretion rate. In addition, the magnetic
configuration of the stellar field after reconnection is not
favorable for formation of the funnel stream. Both factors lead to
quasi-steady accretion through a weak funnel stream, and to lower
accretion rates compared with the case of parallel fields. In the
test run with much higher accretion rate ($\alpha_m\approx 0.1$),
the effect of stress diminishing and other factors becomes
unimportant, and matter accretes in bursty regime like in case
(1).

\item MRI-driven turbulence provides outward angular transport and
inward flow of matter, and gives a significant viscosity.
However, it does not give magnetic diffusivity at the disc-magnetosphere
boundary. We often have a situation where the Prandtl
number is high, $Pr_m= \nu_{vis}/\nu_{dif} >> 1$, at the disc-magnetosphere boundary. In many cases,
reconnection helps matter penetrate towards deeper layers of
the magnetosphere and acts as an efficient diffusivity. In
other cases, small, but finite numerical diffusivity helps.

\item We point out the importance of the initial interaction of
the disc with the magnetosphere, where the disc comes from some
distance after a period of low accretion. Such a situation is
expected in many astrophysical systems where a star has a
dynamically-important magnetic field. Then the inner disc
compresses the star's magnetic flux and moves towards the star.
Matter, after accumulation, flows towards the star and pushes this
magnetic flux to reconnect. This leads to a strong burst in the
accretion rate. This may also lead to a strong X-ray flare
resulting from the reconnection of the magnetic field.
\smallskip

\item A magnetically-dominated corona forms above and below the
disc and slowly expands in the axial direction. The magnetic field
is tightly wrapped and the azimuthal field is 10-100 times
stronger than the poloidal field. The corona is cold and it
expands due to the magnetic pressure gradient.

\end{itemize}

The simulations presented here are restricted to axisymmetric conditions.
Earlier comparisons of axisymmetric and 3D simulations of
MRI-driven accretion discs show good similarity \citep{hawl00},
and we suggest that our disc is sufficient for
investigation of the disc-magnetosphere interaction. However, the
main restriction is probably in the fact that in the full 3D approach
matter can penetrate the magnetic field due to the
Raleigh-Taylor instability \citep{aron76}. Global 3D simulations
show that for some conditions, the disc matter can penetrate
the magnetosphere of the star, in particular in cases of
small tilts of the dipole field and slowly rotating stars
(\citealt{roma05}; \citealt{kulk08}). Hence, it is important to
investigate the MRI-driven accretion onto magnetized stars in full
3D approach.
 We discuss results of global 3D simulations in
\citet{roma11}.

The described research may have multiple applications to different
astrophysical objects. Many  X-ray binaries show accretion in a
flaring regime.  For example, the pulsar SAX J1808.4-3658 shows
flaring activity in the tail of the burst with a quasi-period of 1Hz
\citep{patr09}. We suggest that these flares may be
connected with the process of matter accumulation  at the
disc-magnetosphere boundary with subsequent accretion, similar to
that observed in our case of parallel fields (see Fig.
\ref{push-4}).  In this case the unidirectional magnetic flux
blocks the accretion. This effect is similar to the propeller
effect, where the centrifugal barrier blocks the accretion, and
accretion occurs episodically (e.g.,
\citealt{roma04b,roma05,usty06,roma09}; see also
\citet{spru90,spru93} and \citet{ange10}). In this paper we
considered only slowly rotating stars. Preliminary axisymmetric
simulations of accretion in the propeller regime from MRI-driven
discs show that the interval between accretion events strongly
increases with the spin of the star \citep{usty11}.


\section*{Acknowledgments}

Resources supporting this work were provided by the NASA High-End
Computing (HEC) Program through the NASA Advanced Supercomputing
(NAS) Division at Ames Research Center and the NASA Center for
Computational Sciences (NCCS) at Goddard Space Flight Center. The
authors thank A.A. Blinova, J.F. Hawley and J.M. Stone for helpful
discussions.
   The research was supported by NASA grants
NNX08AH25G, NNX10AF63G and NSF grant AST-0807129.


\appendix

\section{Code description, initial and boundary conditions and reference units}

\subsection{Basic Equations}
\label{sec:app-equations}

The matter flow is described by the equations of ideal MHD:

\begin{equation}\label{eq1}
\displaystyle{ \frac{\partial \rho}{\partial t} + {\bf
\nabla}\cdot \left( \rho {\bf v} \right)} = 0~,
\end{equation}
\begin{equation}\label{eq2}
{\frac{\partial (\rho {\bf v})}{\partial t} + {\bf \nabla}\cdot
{\cal T} } = \rho ~{\bf g}~,
\end{equation}
\begin{equation}\label{eq3}
\frac{\partial {\bf B}}{\partial t} - {\bf \nabla}\times ({\bf v}
\times {\bf B}) = 0~,
\end{equation}
\begin{equation}\label{eq4}
{\frac{\partial (\rho S)}{\partial t} + {\bf \nabla}\cdot ( \rho S
{\bf v} )} =  0~.\
\end{equation}
 Here, $\rho$ is the density and $S$ is the specific entropy; $\bf
v$ is the flow velocity; $\bf B$ is the magnetic field;  $\cal{T}$
is the momentum flux-density tensor; and ${\bf g}$ is the
gravitational acceleration due to the star, which has mass $M$.
The total mass of the disc is negligible compared to $M$.
     The  plasma is considered to be an
ideal gas with adiabatic index $\gamma =5/3$, and $S=\ln(p/
\rho^{\gamma})$. We use cylindrical coordinates $(r, z, \phi)$.
  The condition for
axisymmetry is $\partial /\partial \phi =0$.

\subsection{Reference Units}
\label{sec:app-refunits}

The MHD equations are solved using dimensionless variables
$\widetilde A$. To obtain the physical dimensional values $A$, the
dimensionless values $\widetilde{A}$ should be multiplied by the
corresponding reference units $A_0$; $A=\widetilde{A}A_0$. To
choose the reference units, we first choose the stellar mass
$M_\star$ and radius $R_\star$. The reference units are then
chosen as follows: mass $M_0=M_\star$, distance $R_0=R_\star$,
velocity $v_0=(GM_0/R_0)^{1/2}$, time scale $P_0=2\pi R_0/v_0$,
angular velocity $\Omega_0=v_0/R_0$. The equatorial magnetic field
is determined as $B_\star=\tilde{\mu} B_0$ where $B_0$ is the
reference magnetic field at $R_0$. The parameter  $\tilde\mu$ is
used to vary the magnetic field of the star. We chose $B_0$ such a
way, that for $\tilde\mu=10$, we have a field typical for a given
type of stars. For example, for CTTS, the choice of  $B_0=50$G,
gives equatorial field $B_\star=\tilde\mu=500$G (and polar field
$1000$G) which is typical for CTTSs. Then the models with
larger/smaller values of $\tilde\mu$, correspond to smaller/larger
magnetic field of the star.
 We then define the reference
dipole moment $\mu_{0}=B_0R_0^3$, density $\rho_0=B_0^2/v_0^2$,
pressure $p_0=\rho_0v_0^2$, mass accretion rate
$\dot{M}_0=\rho_0v_0R_0^2$, the torque $N_0=\rho_0v_0^2R_0^3$.
energy per unit time $\dot{E}_0=\rho_0v_0^3R_0^2$.
Temperature $T_0=\mathcal{R}p_0/\rho_0$, where $\mathcal{R}$ is
the gas constant. , and the effective blackbody temperature

Therefore, the dimensionless variables are $\widetilde{r}=r/R_0$,
$\widetilde{v}=v/v_0$, $\widetilde{t}=t/P_0$,
$\widetilde{B}=B/B_0$, $\widetilde{\mu} = \mu/\mu_0$  and so on.
In the subsequent sections, we show dimensionless values for all
quantities and drop the tildes ($\sim$). Our dimensionless
simulations are applicable to different astrophysical objects with
different scales. We list the reference values for typical CTTSs,
cataclysmic variables, and millisecond pulsars in Tab.
\ref{tab:refval}.

\begin{table}
\centering
\begin{tabular}{l@{\extracolsep{0.2em}}l@{}lll}

\hline
& CTTSs       & White dwarfs          & Neutron stars           \\
\hline

{$M_\star(M_\odot)$}              & 0.8              & 1                     & 1.4                     \\
{$R_\star$}                       & $2R_\odot$       & 5000 km               & 10 km                   \\
{$B_{1\star}$ (G)}                & $5\e2$           & $5\e5$                  & $5\e8$                  \\
{$R_0$ (cm)}                      & $1.4\e{11}$      & $1.4\e9$              & $2.9\e6$                \\
{$v_0$ (cm s$^{-1}$)}             & $2.8\e7$         & $5.2\e8$              & $1.4\e{10}$                \\
{$\Omega_0$ (s$^{-1}$)}           & $2\e{-4}$        & 1.03                  & $1.4\e4$                \\
{$P_0$}                           & $0.37$ days      & 6.1 s                 & 0.46 ms                  \\
{$B_0$ (G)}                       & 50               & $5.0\e4$              & $5\e7$                \\
{$\rho_0$ (g cm$^{-3}$)}          & $3.3\e{-12}$     & $3.7\e{-8}$           & $5.3\e{-5}$             \\
{$p_0$ (dy cm$^{-2}$)}            & $2.5\e{3}$       & $1.0\e{10}$           & $1.0\e{16}$             \\
{$\dot M_0(M_\odot$yr$^{-1})$}    & $5.6\e{-8}$      & $1.5\e{-7}$           & $2.3\e{-8}$             \\
{$N_0$ (g cm$^2$s$^{-2}$)}        & $6.9\e{36}$      &   $1.2\e{36}$         & $1.0\e{34}$             \\
{$\dot E_0$ (erg s$^{-1}$)}       & $1.3\e{33}$      & $1.3\e{36}$           & $1.4\e{38}$             \\
\hline
\end{tabular}
\caption{Sample reference units for typical CTTSs, white dwarfs,
and neutron stars. The field $B_{\star}$ corresponds to the case $\tilde\mu=10$. Real dimensional values for variables can be
obtained by multiplying the dimensionless values of variables by
these reference units.} \label{tab:refval}
\end{table}

\subsection{Code description}
\label{sec:app-code-description}

 We solve the full set of
axisymmetric ideal MHD equations in cylindrical coordinates using
the Godunov-type numerical method.   To solve the set of ideal MHD
equations, we use an approximate multi-state HLL Riemann solver
similar to one described recently by \citet{Miyo05}. Compared with
\citet{Miyo05},  we solve an equation for the entropy instead of
full energy equation. This approximation is valid in cases where
shocks are not important, like in our case. In \citet{Miyo05}, the
procedure of the calculations of the MHD equations is described in
detail. Here, we briefly summarize the main features of our
method: the MHD-variables are calculated in four states bounded by
five MHD discontinuities: the contact discontinuity, two Alfven
waves and two fast magnetosponic waves. Compared with
\citet{Miyo05}, we performed the procedure of correction for the
fast magnetosonic waves velocity so that to support the gap
between these waves and the Alfv\'en waves which propagate behind
the fast magnetosonic waves. This helps us to escape the
``non-physical" solutions of the Riemann problem. To satisfy the
condition of the zero divergency of the magnetic field, we
introduced the $\phi-$component of the magnetic field potential
which has been calculated using the method proposed by
\citet{gard05}. To decrease the error in the approximation of the
Lorentz force, we use the splitting of the field to the dipole and
calculated components, $B=B_{dip} + B'$, and dropped terms of the
order of $B_{dip}^2$ which give zero input into the Maxwellian
stress tensor \citep{tana94}. No viscosity or diffusivity has been
added to the code, and hence we investigated accretion driven only
by the MRI-turbulence. Our code has been extensively  tested (see
tests and other details in \citealt{kold11}).

\subsection{Initial conditions}
\label{sec:app-initial-cond}

    The initial conditions  are similar to those
used in \citet{roma05} and \citet{usty06}.   Here, we summarize
these conditions.

The disc is cold with temperature $T_d$ and dense with density
$\rho_d$, while the corona is hot $T_c=1000 T_d$ and rarefied
$\rho_c=0.001\rho_d$. We place both the disc and the corona into
the simulation region. We assume that the initial flow is
barotropic with $\rho=\rho(p)$, and that there is no pressure jump
at the boundary between the disc and corona.
  Then the initial density distribution (in dimensionless units) is the following:
$$
\rho (p)= \left \{
      \begin{array}{lcl}
        p/{\cal R} T_d~,~~~ p>p_b~~~~ {\rm and}~~~~ r \geq
r_b~, \\[0.3cm]
        p/{\cal R} T_c~,~~~~ p<p_b ~~~~{\rm or}~~~~~~r \leq r_b~,
      \end{array}
\right.
$$
\noindent where $p_b$ is the pressure on the surface which
separates the cold matter of the disc from the hot matter of the
corona. On this surface the density jumps from $p_b/T_d$ to
$p_b/T_c$. Here $r_b$ is the inner disc radius.
      Because the density distribution is barotropic,
the angular velocity is constant on coaxial cylindrical surfaces
about the $z-$axis.
        Consequently,
the pressure distribution may be determined from the Bernoulli
equation,
$$
F(p) + \Phi + \Phi_c =E = {\rm const}~.
$$
Here, $\Phi = -GM /|{\bf r}|$  is gravitational potential, $\Phi_c
 = \int_{r \sin \theta}^\infty \Omega^{2} (\xi)\xi
d\xi$ is centrifugal potential, which depends only on the
cylindrical radius $ r\sin \theta$, and
$$
F(p)=\left \{
      \begin{array}{lcl}
        {\cal R} T_d \ln( p/p_b )~,~~~ p>p_b~~~~{\rm and}~~~r \sin \theta >r_b~,
\\[0.3cm]
        {\cal R} T_c \ln (p/p_b) ~,~~~~ p<p_b~~~~{\rm or}~~~~r\sin \theta <r_b~.
      \end{array}
\right.
$$
The angular velocity of the disc is slightly super-Keplerian,
$\Omega(z=0)=\kappa \Omega_K$ ($\kappa = 1+0.02$), due to which
the density and pressure increase towards the periphery.
      Inside the cylinder
$r\leq r_b$ the matter rotates rigidly with angular velocity of
the star $\Omega(r_b)=\kappa (GM /r_b^{3})^{1/2}$. We place the
inner edge of the disc at $r_b=10$ and rotate a star with angular
velocity
      $\Omega(r_b) = 10^{-3/2} \approx 0.03$. The disc evolves,
      but we keep a star rotating slowly to be sure that it
      represents the gravitational well for the incoming matter
      (see the case of rapidly rotating stars in \citealt{usty11}).

Initially, the region is threaded with the dipole magnetic field
of the star. We use different values of the magnetospheric
parameter $\tilde\mu$ from 30 to 2 (see Tab. \ref{tab:models}).
This parameter is responsible for the size of the magnetosphere in
dimensionless runs.
in the disc.
The initial density in the disc varies from
$\rho_d=1$ (at the inner edge) up to $\rho_d=4.6$ (at the outer
boundary).

\subsection{Boundary Conditions}
\label{sec:bound-conditions}

{\it On the star:} The dipole magnetic field is frozen into the
surface of the star, that is the normal to the surface component
of the field  is fixed, while azimuthal component has ``free"
boundary condition, $\partial{B_\phi}/\partial n = 0$. There is
also free boundary conditions for all other variables A, $\partial
A/\partial n=0$. Additional condition is applied to the poloidal
velocity which forbids matter to flow out of the star. In
addition, in the last grid we adjust the total velocity vector to
be parallel to the total magnetic field vector. This helps to
strengthen the frozen-in condition near the star and to enhance
the divergency-free condition at the boundary.~ {\it The top and
bottom boundaries:} The values of density $\rho$, entropy $s$ and
azimuthal field $B_\phi$ are fixed. Free conditions for all other
variables ($B_r$, $B_z$, $v_r$, $v_z$, $v_\phi$). In addition,
$v_z=0$ in case if matter flows into the simulation region. ~{\it
The side boundary:} The side boundary is divided into the ``the
disc region" at $|z| < z_d$ and ``corona region" at $|z| > z_d$,
where
$$
z_d=\sqrt{\bigg[\frac{GM}{F_c(R_{out})}\bigg]^2 - R_{out}^2},
$$
where $F_c(R_{out})=a GM/R_{out}$ is centrifugal potential and
$R_{out}$ is the external radius. In the disc ($|z| < z_d$) we
have an inflow condition for the matter. Namely, we push matter
into the region with small velocity:
$$
v_r=-\delta \frac{3}{2}\frac{p}{\rho v_K(R_{out})}, ~~\delta=0.02,
$$
and with poloidal magnetic field corresponding to the initial
magnetic field at $r=R_{out}$. We also fix $\rho$, $s$ and
$B_\phi$, while condition for $v_z$ is free condition. In the
corona, $|z| < z_d$ conditions are similar to those on top and
bottom boundaries.


\begin{thebibliography}{}



\bibitem[\protect\citeauthoryear{Arons \& Lea}{1976}]{aron76} Arons, J., Lea, S. M., 1976, ApJ,
207, 914
\bibitem[\protect\citeauthoryear{Balbus \& Hawley}{1991}]{balb91} Balbus, S.A., \& Hawley, J.F. 1991, ApJ, 376, 214
\bibitem[\protect\citeauthoryear{Balbus \& Hawley}{1998}]{balb98} Balbus, S.A., \& Hawley, J.F. 1998,  Reviews of Modern Physics, Volume 70, 1
\bibitem[\protect\citeauthoryear{Beckwith, et al.}{2009}]{beck09} Beckwith, K., Hawley, J.F., \& Krolik, J.H. 2009,
ApJ, 707, 428
\bibitem[\protect\citeauthoryear{Bessolaz et al.}{2008}]{bess08} Bessolaz N., Zanni C., Ferreira J., Keppens R., Bouvier J., 2008, A\&A, 478, 155
\bibitem[\protect\citeauthoryear{Bouvier et al.}{2005}]{bouv05} Bouvier, J., Alencar, S., Harries, T.J.,  Johns-Krull, C.M.,
\& Romanova, M.M. 2005, Protostars and Planets V {\it
``Magnetospheric Accretion in Classical T Tauri Stars"}

\bibitem[\protect\citeauthoryear{D'Angelo \& Spruit}{2010}]{ange10}     D'Angelo, C. R., Spruit, H. C. 2010, MNRAS,
406, 1208

\bibitem[\protect\citeauthoryear{Frey et al}{2003}]{frey03} Frey, H. U., Phan, T. D., Fuselier, S. A., Mende, S. B. 2003, Nature, 426, 533

\bibitem[\protect\citeauthoryear{Gammie \& Menou}{1998}]{gamm98} Gammie, C.F., Menou, K. 1998, ApJ, 492,
L75
\bibitem[\protect\citeauthoryear{Gardiner \& Stone}{2005}]{gard05}  Gardiner, T.A., Stone, J.M.
2005, J. Comp. Phys., 205, 509
\bibitem[\protect\citeauthoryear{Goodman \& Xu}{1994}]{good94} Goodman, J., \& Xu, G. 1994, ApJ, 432,
213
\bibitem[\protect\citeauthoryear{Goodson et al.}{1997}]{good97} Goodson, A.P., Winglee, R. M., \& B\"ohm, K.-H. 1997, ApJ,
489, 199
\bibitem[\protect\citeauthoryear{Goodson et al.}{1999}]{good99} Goodson, A.P., B\"ohm, K.-H., Winglee, R. M. 1999, ApJ, 524,
142
\bibitem[\protect\citeauthoryear{Hawley et al.}{1995}]{hawl95} Hawley, J.F,, Gammie, C.F., Balbus, S.A. 1995, ApJ, 440,
742
\bibitem[\protect\citeauthoryear{Hawley}{2000}]{hawl00} Hawley, J.F, 2000, ApJ, 528, 462
\bibitem[\protect\citeauthoryear{Hawley et al.}{2001}]{hawl01} Hawley, J.F, Balbus, S.A., \& Stone, J.M. 2001, ApJ Letters,
554, L49
\bibitem[\protect\citeauthoryear{Hawley \& Krolik}{2002}]{hawl02} Hawley, J.F, \& Krolik, J.H. 2002, ApJ 566, 164
\bibitem[\protect\citeauthoryear{Koldoba et al.}{2011}]{kold11} Koldoba A. V., Ustyugova G. V., Romanova M. M., Lovelace R. V. E. 2011, in prep.
\bibitem[\protect\citeauthoryear{Kulkarni \& Romanova}{2005}]{kulk05} Kulkarni A. K., Romanova M. M. 2005, ApJ, 633, 349
\bibitem[\protect\citeauthoryear{Kulkarni \& Romanova}{2008}]{kulk08} Kulkarni, A. K., Romanova, M. M. 2008, MNRAS, 386, 673
\bibitem[\protect\citeauthoryear{Long et al.}{2005}]{long05} Long M., Romanova M. M., Lovelace R. V. E. 2005, ApJ, 634, 1214
\bibitem[\protect\citeauthoryear{Latter et al.}{2009}]{latt09} Latter, H. N., Lesaffre, P., \& Bablbus, S. A. 2009, MNRAS,
394, 715
\bibitem[\protect\citeauthoryear{Long et al.}{2008}]{long08} ---------. 2008, MNRAS, 386, 1274

\bibitem[\protect\citeauthoryear{Lovelace et al.}{1995}]{love95} Lovelace, R. V. E.,
 Romanova, M. M., Bisnovatyi-Kogan, G. S. 1995, MNRAS, 275, 244

\bibitem[\protect\citeauthoryear{Miller \& Stone}{1997}]{mill97}
Miller, K.A., \& Stone, J.M. 1997, ApJ, 489, 890
\bibitem[\protect\citeauthoryear{Miller \& Stone}{2000}]{mill00}
Miller, K.A., \& Stone, J.M. 2000, ApJ, 534, 398
\bibitem[\protect\citeauthoryear{Miyoshi \& Kusano}{2005}]{Miyo05} Miyoshi, T., Kusano, K. 2005, J. Comp. Phys., 208,
315
\bibitem[\protect\citeauthoryear{Patruno et al.}{2009}]{patr09} Patruno, A., Watts, A., Klein Wolt, M., Wijnands, R., van der
Klis, M. 2009, ApJ, 707, 1296
\bibitem[\protect\citeauthoryear{Romanova et al.}{2008}]{roma08} Romanova M. M., Kulkarni A. K., Lovelace R. V. E., 2008, ApJ, 673, L171
\bibitem[\protect\citeauthoryear{Romanova et al.}{2002}]{roma02} Romanova M. M., Ustyugova G. V., Koldoba A. V., Lovelace R. V. E., 2002, ApJ, 578, 420
\bibitem[\protect\citeauthoryear{Romanova et al.}{2004}]{roma04a} ---------. 2004a, ApJ, 610, 920
\bibitem[\protect\citeauthoryear{Romanova et al.}{2004}]{roma04b} Romanova M. M., Ustyugova G. V., Koldoba A. V., Lovelace R. V.
E., 2004b, ApJ, 616, L151
\bibitem[\protect\citeauthoryear{Romanova et al.}{2005}]{roma05} Romanova M. M., Ustyugova G. V., Koldoba A. V., Lovelace R. V.
E., 2005, ApJ, 635, L165
\bibitem[\protect\citeauthoryear{Romanova et al.}{2009}]{roma09} ---------. 2009, MNRAS, 399, 1802
\bibitem[\protect\citeauthoryear{Romanova et al.}{2011}]{roma11} ---------. 2011,
MNRAS, in preparation
\bibitem[\protect\citeauthoryear{Romanova et al.}{2003}]{roma03} Romanova M. M., Ustyugova G. V., Koldoba A. V., Wick J. V., Lovelace R. V. E., 2003, ApJ, 595, 1009
\bibitem[\protect\citeauthoryear{Shakura \& Sunyaev}{1973}]{shak73}
Shakura, N.I., \& Sunyaev, R.A. 1973, A\&A, 24, 337
\bibitem[\protect\citeauthoryear{Spruit \& Taam}{1990}]{spru90} Spruit, H.C., Taam,
R.E., 1990, ApJ, 229, 475
\bibitem[\protect\citeauthoryear{Spruit \& Taam}{1993}]{spru93} Spruit, H.C., Taam,
R.E., 1993, ApJ, 402, 593

\bibitem[\protect\citeauthoryear{Stone et al.}{1996}]{ston96} Stone, J.M., Hawley, J.F., Balbus, S.A., Gammie, C.F.
1996, ApJ, 463, 656
\bibitem[\protect\citeauthoryear{Stone \& Pringle}{2001}]{ston01} Stone, J.M.,
\& Pringle, J.E. 2001, MNRAS, 322, 461
\bibitem[\protect\citeauthoryear{Tanaka}{1994}]{tana94} Tanaka, T. 1994, J. Comp. Phys., 111, 381
\bibitem[\protect\citeauthoryear{Ustyugova et al.}{2006}]{usty06} Ustyugova, G.V., Koldoba, A.V., Romanova, M.M., \& Lovelace,
R.V.E. 2006, ApJ, 646, 304
\bibitem[\protect\citeauthoryear{Ustyugova et al.}{2011}]{usty11} Ustyugova, G.V., Koldoba, A.V., , Romanova, M.M. \& Lovelace,
R.V.E. 2011, in preparation
\bibitem[\protect\citeauthoryear{Van der Klis}{2000}]{vander00} Van der Klis, M.  2000, Ann. Rev. Astron. Astrophys., 38,
717, {\it ``Millisecond Oscillations in X-ray Binaries"}

\bibitem[\protect\citeauthoryear{Warner}{2004}]{warn04} Warner, B.  2004, PASP, Volume 116, Issue 816, pp.115, {\it
``Rapid Oscillations in Cataclysmic Variables"}


\end{thebibliography}
\end{document}